\documentclass[pdflatex,sn-basic]{sn-jnl}

\usepackage{graphicx}%
\usepackage{multirow}%
\usepackage{amsmath,amssymb,amsfonts}%
\usepackage{amsthm}%
\usepackage{mathrsfs}%
\usepackage[title]{appendix}%
\usepackage{xcolor}%
\usepackage{textcomp}%
\usepackage{manyfoot}%
\usepackage{booktabs}%
\usepackage{algorithm}%
\usepackage{algorithmicx}%
\usepackage{algpseudocode}%
\usepackage{listings}%
\usepackage{sidecap}
\usepackage{ulem}
\let\orcidlogo\undefined \usepackage{orcidlink}

\renewcommand{\vec}[1]{\boldsymbol{#1}}

\newcommand\bb[1]{\mbox{\boldmath{$#1$}}}
\newcommand\grad{\bb{\nabla}}
\newcommand\bcdot{\,\bb{\cdot}\,}

\newcommand\btimes{\,\bb{\times}\,}

\let\emph\relax
\DeclareTextFontCommand{\emph}{\itshape}

\begin{document}

\title[Electrons in tangled fields]{Transport of electrons in tangled magnetic fields
}


\author*[1]{\fnm{Daniel} \sur{Verscharen}\orcidlink{0000-0002-0497-1096}}\email{d.verscharen@ucl.ac.uk}

\author[2]{\fnm{Natasha} \sur{Jeffrey} \orcidlink{0000-0001-6583-1989}}\email{natasha.jeffrey@northumbria.ac.uk}

\author[3]{\fnm{Anton} \sur{Artemyev} \orcidlink{0000-0001-8823-4474}}\email{aartemyev@igpp.ucla.edu}

\author[1]{\fnm{Jesse~T.} \sur{Coburn}\orcidlink{0000-0002-2576-0992}}\email{j.coburn@imperial.ac.uk}\equalcont{Now at: The Blackett Laboratory, Imperial College London, London SW7 2AZ, United Kingdom.}

\author[4,5]{\fnm{Matthew~W.} \sur{Kunz} \orcidlink{0000-0003-1676-6126}}\email{mkunz@princeton.edu}

\author[6]{\fnm{Oreste} \sur{Pezzi} \orcidlink{0000-0002-7638-1706}}\email{oreste.pezzi@istp.cnr.it} 

\author[7]{\fnm{Mario} \sur{Riquelme} \orcidlink{0000-0003-2928-6412}}\email{marioriquelme@uchile.cl} 

\author[8,9,10]{\fnm{Ida} \sur{Svenningsson} \orcidlink{0000-0003-1469-1116}}\email{ida.svenningsson@chalmers.se}

\author[11]{\fnm{Lynn~B.} \sur{Wilson III}\orcidlink{0000-0002-4313-1970}}\email{lynn.b.wilson@nasa.gov}

\affil[1]{\orgdiv{Mullard Space Science Laboratory}, \orgname{University College London}, \orgaddress{\street{Holmbury House}, \city{Dorking}, \postcode{RH5~6NT}, \country{United Kingdom}}}

\affil[2]{%
\orgdiv{Department of Mathematics, Physics and Electrical Engineering}, 
\orgname{Northumbria University}, 
\orgaddress{
\city{Newcastle upon Tyne}, \postcode{NE1 8ST}, \country{United Kingdom}}}

\affil[3]{%
\orgdiv{Department of Earth, Planetary, and Space Sciences}, 
\orgname{University of California}, 
\orgaddress{
\city{Los Angeles, CA}, \postcode{90095}, \country{USA}}}

\affil[4]{%
\orgdiv{Department of Astrophysical Sciences},
\orgname{Princeton University},
\orgaddress{\city{Princeton, NJ}, 
\postcode{08544}, \country{USA}}}

\affil[5]{%
\orgname{Princeton Plasma Physics Laboratory},
\orgaddress{\city{Princeton, NJ}, \postcode{08543}, \country{USA}}}

\affil[6]{\orgdiv{Institute for Plasma Science and Technology}, \orgname{National Research Council of Italy}, \orgaddress{\street{G. Amendola 122/D}, \city{Bari}, \postcode{I-70126}, \country{Italy}}}

\affil[7]{\orgdiv{Departamento de F\'isica},
\orgname{Facultad de Ciencias F\'isicas y Matem\'aticas (FCFM), Universidad de Chile},
\orgaddress{\street{Beauchef 850}, \city{Santiago}, \country{Chile}}}

\affil[8]{
\orgname{Swedish Institute of Space Physics}, 
\orgaddress{
\city{Uppsala}, \postcode{751 21}, \country{Sweden}}}

\affil[9]{%
\orgdiv{Department of Physics and Astronomy}, 
\orgname{Uppsala University}, 
\orgaddress{
\city{Uppsala}, \postcode{751 20}, \country{Sweden}}}

\affil[10]{%
\orgdiv{Department of Physics}, 
\orgname{Chalmers University of Technology}, 
\orgaddress{
\city{Gothenburg}, \postcode{412 96}, \country{Sweden}}}

\affil[11]{\orgdiv{NASA Goddard Space Flight Center}, \orgname{Heliophysics Division}, \orgaddress{\street{Code 672, Bldg. 21, Rm. 143A}, \city{Greenbelt, MD}, \postcode{20771}, \country{United States}}}


\abstract{Cosmic magnetic fields are typically inhomogeneous and often highly tangled due to large-scale plasma flows, turbulence, and instabilities. If the variations in the magnetic field occur on scales that are large compared to the gyro-radius of the plasma electrons, the electrons are primarily confined to gyro-centre trajectories along the field lines. Therefore, in-situ electron measurements help us map out the connectivity of the magnetic field in space plasmas. Gyro-centre drifts, wave--particle interactions, trapping, and cross-field diffusion are processes related to field inhomogeneities and fluctuations; they have the potential to modify or even disrupt the transport of electrons along field lines. We introduce the basic principles of electron transport in tangled magnetic fields and review the creation of tangled fields through turbulence and instabilities as well as the modulation of parallel electron transport through kinetic instabilities. We then describe trapping and de-trapping effects in inhomogeneous magnetic fields, as well as electron diffusion and energisation across the magnetic field. The transport of electrons in  tangled fields results from a complex interplay of plasma processes that occur on a broad range of scales. A combination of in-situ plasma measurements, remote-sensing plasma observations, and plasma theory and simulations is required to resolve this contemporary challenge to the fields of heliophysics and astrophysics.}




\maketitle


\section{Introduction}\label{sect:intro}

In a homogeneous, static magnetic field $\vec B$, the Lorentz force deflects electrons in the direction perpendicular to their velocity vector $\vec v$ and to $\vec B$. Consequently, the electrons undergo gyration at their cyclotron frequency 
\begin{equation}\label{gyrofreq}
\Omega_{\mathrm e}=\frac{eB}{m_{\mathrm e}c},
\end{equation}
where $e<0$ is the charge of an electron, $m_{\mathrm e}$ is the mass of an electron, and $c$ is the speed of light.\footnote{Equation~(\ref{gyrofreq}) also applies in the relativistic case as long as $m_{\mathrm e}$ accounts for the relativisic correction  $m_{\mathrm e}=\gamma m_{0\mathrm e}$, where $\gamma=1/\sqrt{1-v^2/c^2}$ is the Lorentz factor and $m_{0\mathrm e}$ is the rest mass of an electron. In our convention, $\Omega_{\mathrm e}<0$.} In the absence of an electric field, the particles conserve their kinetic energy and spiral about the magnetic field $\vec B$. In a reference frame in which the centre of this spiral motion is at rest, we denote the magnitude of the velocity vector in the plane perpendicular to $\vec B$ as $v_{\perp}$ and the component of the velocity vector parallel to $\vec B$ as $v_{\parallel}$. If the field is homogeneous and no other forces act on the electron, $v_{\perp}$ and $v_{\parallel}$ are constant. The radius of the spiral trajectory is given by the gyro-radius
\begin{equation}
\rho_{\mathrm e}=\frac{v_{\perp}}{|\Omega_{\mathrm e}|}.
\end{equation}
In most space and astrophysical plasmas, $\rho_{\mathrm e}$ is significantly smaller than the characteristic scales of the system \citep{verscharen19}, so that gyration confines electrons into a cylindrical volume with a small radius that is centred on a given magnetic-field line and the electrons are \emph{magnetised}. While the radius of the accessible cylindrical volume is small, the electrons are free to move parallel to $\vec B$. For this reason, most electron transport occurs along the magnetic field.

All space and astrophysical plasma systems ultimately break the assumptions of a homogeneous and static magnetic field though. Large-scale flows, instabilities, and turbulence tangle magnetic-field lines, leading to complex field geometries \citep{bruno13,laitinen23}. Even though electron transport still occurs mostly along the magnetic field, additional effects modify the electron motion compared to the homogeneous case. Since most tangling processes occur on timescales that are large compared to the associated electron gyro-motion (i.e., $\gg 1/|\Omega_{\mathrm e}|$), we focus on spatial inhomogeneity in this article while assuming that the field is quasi-static in time.

\begin{figure}[ht]
\centering
\includegraphics[width=\textwidth]{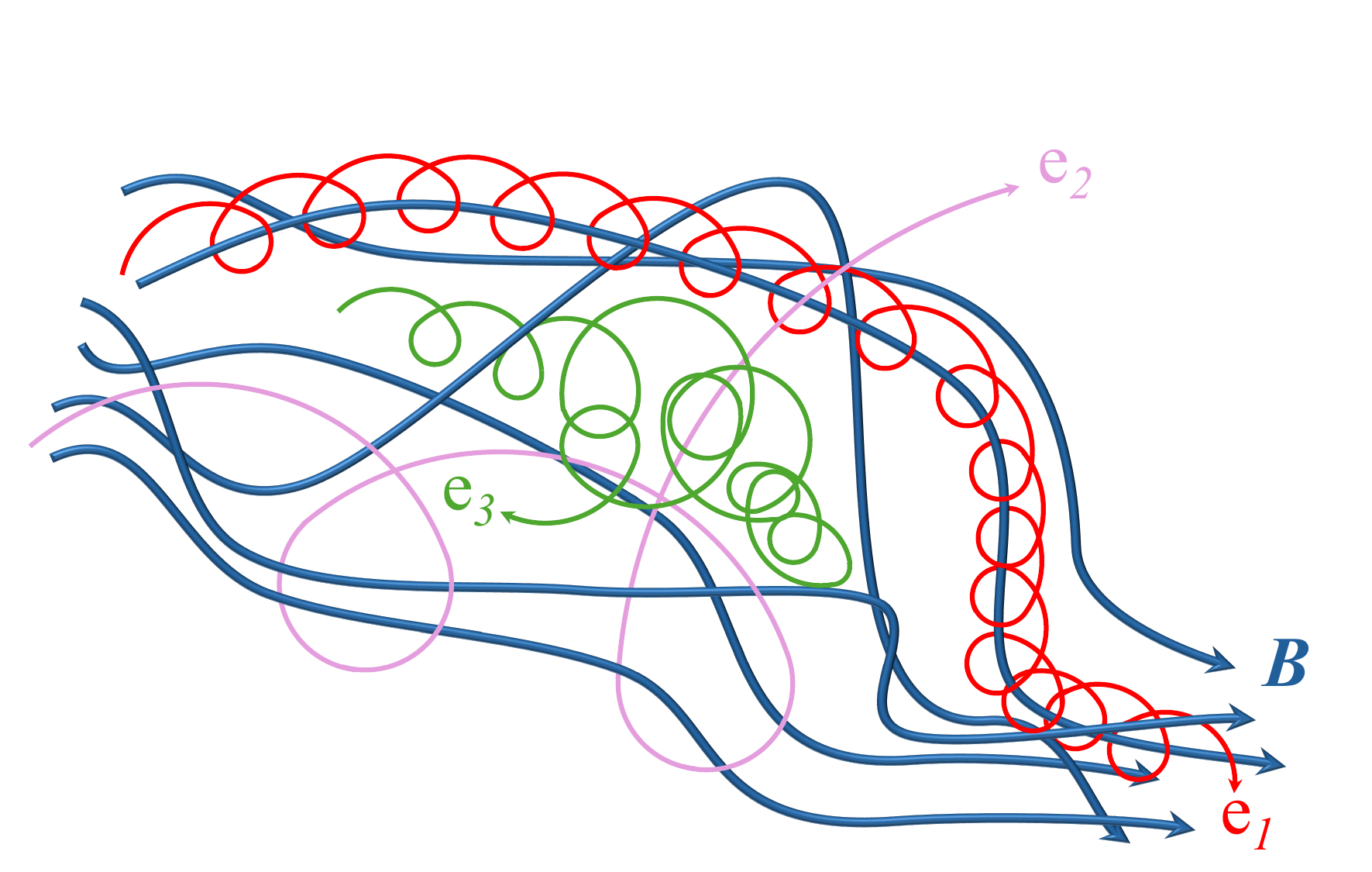}
\caption{Example trajectories of electrons in tangled magnetic-field lines ($\vec B$, blue). Electron~1 (e$_1$, red) follows a given field line. Electron~2 (e$_2$, purple) diffuses across field lines. Electron~3 (e$_3$, green) undergoes reflection at a mirror point }\label{tangled_fields_sketch}
\end{figure}

Figure~\ref{tangled_fields_sketch} illustrates three example electron trajectories in tangled magnetic fields. The field lines are indicated in blue. 
Electron~1 (e$_1$, red) has a small $\rho_{\mathrm e}$ compared to the scales over which $\vec B$ changes. Therefore, e$_1$ gyrates about a given field line and follows this field line when averaged over the electron's gyro-motion. The path length of the gyro-centre trajectory of a magnetised electron between two given points is longer if the field line is tangled rather than straight between both points. 

The magnetic moment 
\begin{equation}
\mu=\frac{m_{\mathrm e}v_{\perp}^2}{2B}
\end{equation}
of e$_1$ remains constant along the trajectory of the particle. As long as the electric field is negligible along the electron trajectory,  the electron kinetic energy
\begin{equation}
W=\frac{1}{2}m_{\mathrm e}\left(v_{\perp}^2+v_{\parallel}^2\right)
\end{equation}
also remains constant.
In the presence of an electric field $\vec E$ with a component perpendicular to the magnetic field or a large-scale (compared to $\rho_{\mathrm e}$) inhomogeneity of the magnetic field, the gyro-centre of the gyrating electron follows the guiding-centre motion with the velocity \citep{hazeltine73,chen16}
\begin{equation}\label{drifts}
\vec V_{\mathrm g}=c\frac{\vec E\btimes \vec B}{B^2}+\frac{\mu}{m_{\mathrm e}\Omega_{\mathrm e}}\hat{\vec b}\btimes \grad B+\frac{v_{\parallel}^2}{\Omega_{\mathrm e}}\hat{\vec b}\btimes\left(\hat{\vec b}\bcdot \grad\right)\hat{\vec b}+v_{\parallel}\hat{\vec b},
\end{equation}
where $\hat{\vec b}=\vec B/B$. The magnetic moment $\mu$ is here measured in a reference frame that drifts with the $\vec E\btimes \vec B$ drift, leading to the definition 
\begin{equation}
v_{\perp}^2=\left(\vec v-v_{\parallel}\hat{\vec b}-c\frac{\vec E\btimes\vec B}{B^2}\right)^2,
\end{equation}
where $v_{\parallel}=\vec v\bcdot  \hat{\vec b}$.
In Eq.~(\ref{drifts}), the first term describes the $\vec E\btimes \vec B$ drift. In a system in which the frozen-in condition holds, the field lines are advected at the speed $c\vec E\btimes \vec B/B^2$ (see also Sect.~\ref{turb_cascade}), so that the $\vec E\btimes \vec B$ drift typically conserves the association of a gyrating electron with its magnetic-field line. 
The second term in Eq.~(\ref{drifts}) describes the gradient drift. It results from a spatial variation in the gyro-radius of the electron. When during its gyro-motion the electron is exposed to spatial variation in the magnitude of $B$, its gyro-radius is smaller (at times of larger $B$) or larger (at times of smaller $B$). This variation causes the electron to drift in the direction of $-\hat{\vec b}\btimes \grad B$, given that $\Omega_{\mathrm e}<0$. The third term in Eq.~(\ref{drifts}) describes the curvature drift. It results from the motion of the electron along a curved magnetic-field line. An electron following a curved field line experiences a centrifugal force directed perpendicular to the curved field line, which alternately accelerates and decelerates the gyro-motion. Therefore, an electron drifts in the direction $-\hat{\vec b}\btimes(\hat{\vec b}\bcdot \grad)\hat{\vec b}$, i.e., perpendicular to $\vec B$ and the curvature vector $\vec{\kappa}=(\hat{\vec b}\bcdot\grad)\hat{\vec b}$. The fourth term in Eq.~(\ref{drifts}) describes the motion of the electron along the magnetic field.   
The gradient and curvature drifts always occur in combination and allow electrons to leave their associated magnetic-field lines. They play a key role, for example, in laboratory plasmas in which particles drift off the associated field lines can lead to loss of plasma confinement, unless the field geometry returns particles to the plasma core.

Electron~2 (e$_2$, purple) has a greater $v_{\perp}$ compared to e$_1$. Therefore, its gyro-radius $\rho_{\mathrm e}$ is larger than the gyro-radius of e$_1$ at the same $B$. The electron still undergoes gyration in regions of large $B$ as seen at the beginning of its shown trajectory. Later, it encounters regions with variation in $\vec B$ on scales that are comparable to or smaller than its $\rho_{\mathrm e}$. There, the electron loses its strong magnetisation and moves, while still undergoing some deflections due to the Lorentz force, by some distance across the field. Its magnetic moment $\mu$ is not conserved because the electron encounters a significant change in $\vec B$ during individual gyro-orbits. The particle undergoes spatial diffusion across the magnetic field. In non-static field configurations, a fast time variation of the magnetic field can also contribute to the breaking of $\mu$-conservation.

Electron~3 (e$_3$, green) represents a magnetised particle that travels from a region of low $B$ into a region of high $B$. Due to the conservation of $\mu$ and $W$, the electron's $v_{\perp}$ increases and its $v_{\parallel}$ decreases as $B$ increases. The gyro-centre of the electron propagates to a point where $v_{\parallel}=0$. This point is called the \emph{mirror point}. The gyrating electron undergoes reflection at this point, and its gyro-centre then propagates back into the region of lower $B$. If the gyrating electron encounters another region with higher $B$ after propagating through the location of lowest $B$ along its trajectory, it may undergo consecutive reflections between both mirror points. In this case, we refer to the electron as being \emph{trapped}. We discuss trapping effects in Sect.~\ref{sect:transport}.

Through the interaction of electrons with the tangled magnetic field, electrons couple to all other plasma species electromagnetically. Therefore, all these processes give rise to an anomalous transfer of momentum from electrons to ions and vice versa. In collisionless plasmas, this inter-species transfer of momentum creates \emph{anomalous resistivity} and thus lowers the conductivity of the plasma by reducing the free streaming of electrons \citep{drummond62,davidson75,davidson77,papadopoulos77}. This anomalous resistivity plays a crucial role, for example, in magnetic reconnection \citep{mozer11,le18,graham22} and magnetogenesis \citep{Schekochihin2006,Mogavero2014} in collisionless plasmas.

\begin{figure}[t]
\centering
\includegraphics[width=\textwidth]{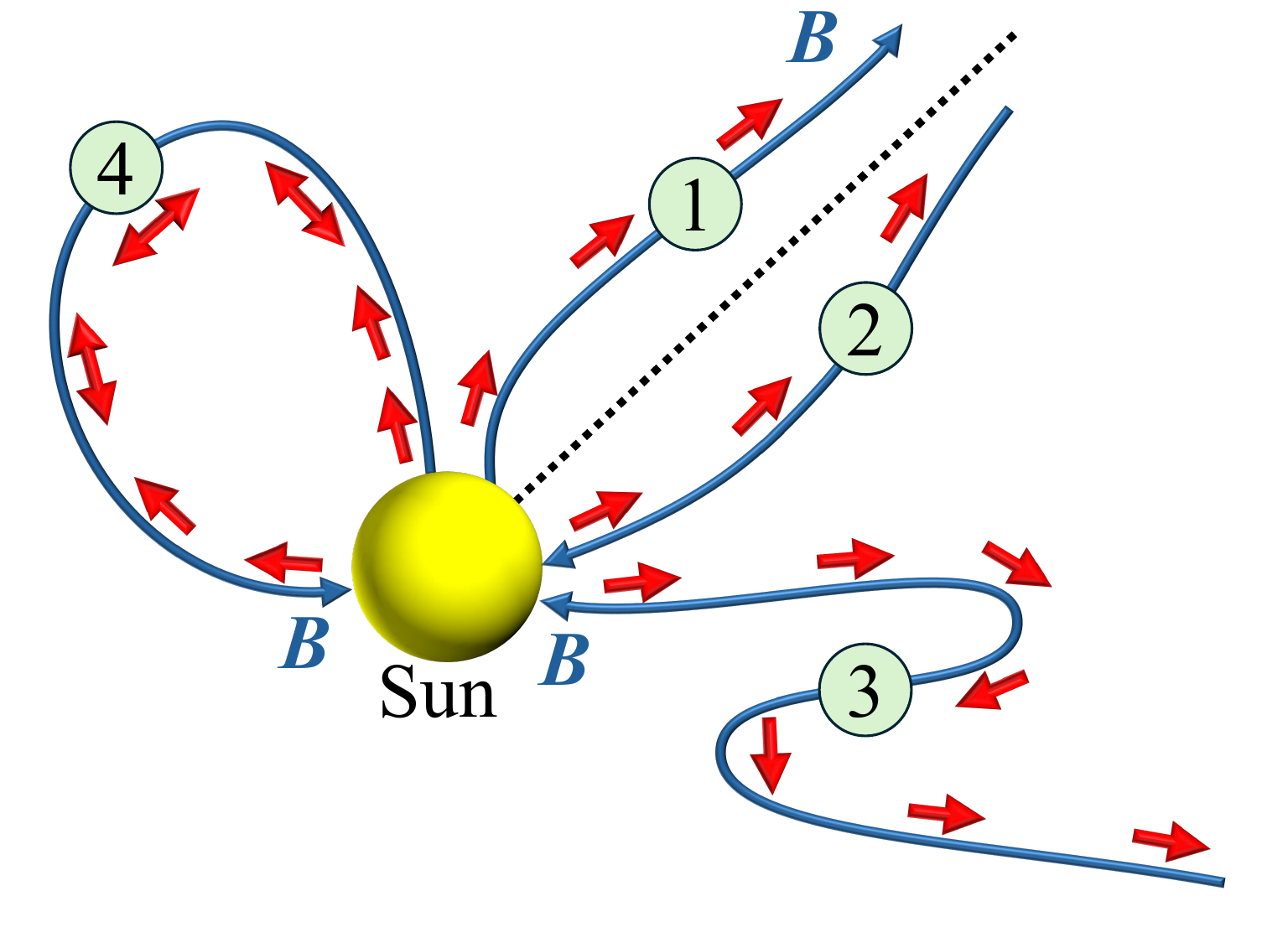}
\caption{The strahl can be used as a tool to measure magnetic connectivity in the solar wind. The blue curves represent magnetic-field lines connected with the Sun. The red arrows indicate the local direction of the heat flux vector. The dashed black line represents a global separation of field-line connectivity (e.g., the heliospheric current sheet). The green circles illustrate in-situ measurement points}\label{connectivity}
\end{figure}
In space plasmas accessible to in-situ observations, we exploit the strong magnetisation of electrons to ascertain the magnetic connectivity of the interplanetary magnetic field with the Sun when measured by a spacecraft. In the solar wind, for instance, the majority of the electron heat flux is visible as a field-aligned beam in the electron velocity distribution function, referred to as the \emph{strahl}  \citep{pilipp87,maksimovic97,pierrard01}. The strahl feature originates in the solar corona and consists of magnetised electrons that move along magnetic-field lines and away from the Sun \citep{owens17}. A local measurement of the strahl pitch-angle distribution thus allows us to determine the connectivity of the local field line to the solar corona \citep{gosling87,crooker04,macneil20}.  Figure~\ref{connectivity} illustrates this approach. At measurement point 1, the strahl propagates into the direction of the field (parallel to $\vec B$). The electron heat-flux vector $\vec q_{\mathrm e}$ is thus likewise parallel to $\vec B$. At measurement point 2, the opposite situation occurs: the strahl propagates into the direction anti-parallel to $\vec B$. When crossing a field-line reversal, such as at measurement point 3, the strahl propagates towards the Sun but at the same orientation with respect to $\vec B$ as before and after the interval when the field reversed. At measurement point 4, the electron pitch-angle distribution shows bi-directional strahl due to the connectivity to  both coronal footpoints of a magnetic-field loop. 
The same technique is also a powerful tool for the analysis of magnetic connectivity in planetary plasma environments \citep[e.g.,][]{frahm2006,tsang2015}.

\begin{figure}[!t]
  \centering
    {\includegraphics[width=0.7\textwidth]{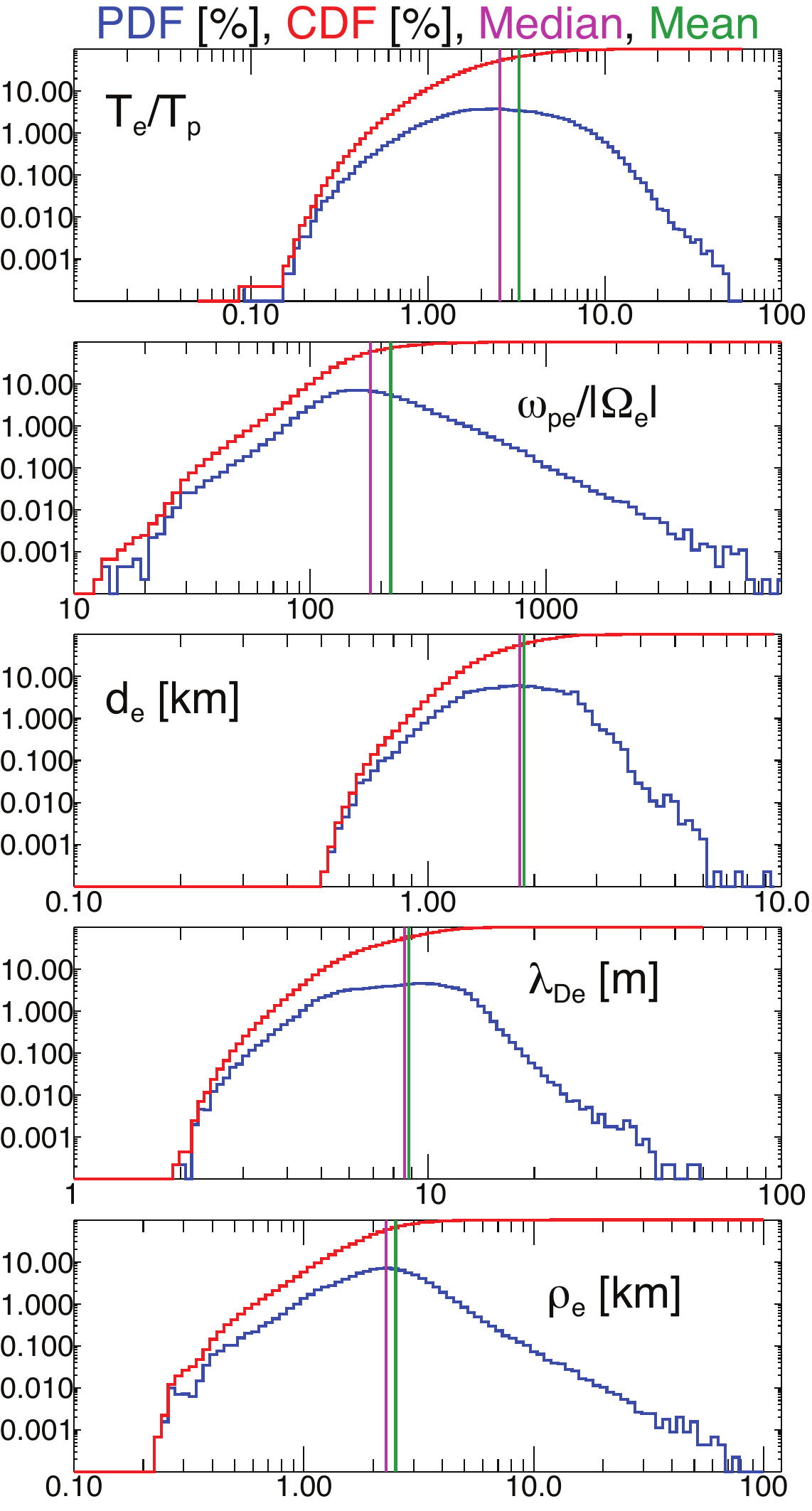}}
    \caption{PDFs (blue histograms) and CDFs (red histograms) of several solar-wind parameters observed by \emph{Wind} near 1~au.  The panels (from top to bottom) are as follows: $T_{\mathrm e}/T_{\mathrm p}$, $\omega_{\mathrm{pe}}/|\Omega_{\mathrm e}|$, $d_{\mathrm e}$, $\lambda_{\mathrm{De}}$, and  $\rho_{\mathrm e}$.  For each distribution, we  show the median (magenta) and mean (green) values of the data as vertical lines. The data are taken from \citet[][]{wilsoniii23a}}
    \label{fig:PDFsCDFsSWParams1AU}
\end{figure}

The solar wind provides us with a unique laboratory to study electron transport in tangled fields. It gives us access to direct in-situ measurements of electron velocity distribution functions and electromagnetic fields. At the same time, the plasma is so variable that, by sampling long statistical datasets, we have access to a broad range of plasma conditions. Figure~\ref{fig:PDFsCDFsSWParams1AU} shows probability density functions (PDFs; blue histograms) and cumulative distribution functions (CDFs; red histograms) of several solar-wind parameters observed by the \emph{Wind} spacecraft at a heliocentric distance of about 1~au.  It is important to remember that the five plasma parameters reported in Figure~\ref{fig:PDFsCDFsSWParams1AU} are statistically correlated. Therefore, although the solar wind provides us with a wide range of plasma conditions, not all parameter combinations can be sampled with the same statistical reliability in spacecraft measurements in the solar wind at 1\,au.

The top panel shows the statistical distribution of the electron-to-proton temperature ratio, $T_{\mathrm e}/T_{\mathrm p}$, indicating that on average $T_{\mathrm e}>T_{\mathrm p}$ in equatorial solar wind, which is a typical feature of slow solar wind \citep{salem2023}. The second panel shows the ratio $\omega_{\mathrm{pe}}/|\Omega_{\mathrm e}|$ between the electron plasma frequency 
\begin{equation}
\omega_{\mathrm{pe}}=\sqrt{\frac{4\pi n_{\mathrm e}e^2}{m_{\mathrm e}}},
\end{equation}
where $n_{\mathrm e}$ is the electron density, and the electron cyclotron frequency.  
The third panel shows the statistical distribution of the electron inertial length 
\begin{equation}
d_{\mathrm e}=\frac{c}{\omega_{\mathrm{pe}}},
\end{equation}
and the fourth panel shows the statistical distribution of the Debye length
\begin{equation}
\lambda_{\mathrm{De}}=\sqrt{\frac{k_{\mathrm B}T_{\mathrm e}}{4\pi n_{\mathrm e}e^2}},
\end{equation}
where $k_{\mathrm B}$ is the Boltzmann constant.
The fifth panel shows the statistical distribution of the electron  gyro-radius $\rho_{\mathrm e}$, where we evaluate $\rho_{\mathrm e}$ as the thermal gyro-radius, i.e., the gyro-radius of an electron with $v_{\perp}=w_{\mathrm e}$,
where
\begin{equation}
w_{\mathrm e}=\sqrt{\frac{2k_{\mathrm B}T_{\mathrm e}}{m_{\mathrm e}}}
\end{equation}
is the thermal speed of the electrons. We note that, with these definitions, the squared ratio of the electron  gyro-radius to the electron inertial length describes the ratio between the electron thermal pressure to the magnetic pressure:
\begin{equation}
\beta_{\mathrm e}=\left(\frac{\rho_{\mathrm e}}{d_{\mathrm e}}\right)^2=\frac{8\pi n_{\mathrm e}k_{\mathrm B}T_{\mathrm e}}{B^2}.
\end{equation}
The mean of $\beta_{\mathrm e}$ in the dataset shown in Fig.~\ref{fig:PDFsCDFsSWParams1AU} is 2.31, while its median is 1.09.

The tangling of the magnetic field and the impact of tangled fields on electron transport depend on the spatial and temporal scales at which the fluctuations in the field occur. For the context of electron transport, the comparison of the spatial scales of the field's variation with $d_{\mathrm e}$ and $\rho_{\mathrm e}$ (see also Fig.~\ref{tangled_fields_sketch}) provides valuable insights. For context, the mean $d_{\mathrm e}$ is 1.87\,km, the median $d_{\mathrm e}$ is 1.82\,km, the mean $\lambda_{\mathrm{De}}$ is 8.84\,m, the median $\lambda_{\mathrm{De}}$ is 8.58\,m, the mean $\rho_{\mathrm e}$ is 2.51\,km, and the median $\rho_{\mathrm e}$ is 2.28\,km. 
Figure~\ref{fig:PDFsCDFsSWParams1AU} also highlights that the near-Earth solar wind exhibits a large separation of the characteristic electron scales, which poses big challenges when attempting to model this system numerically with kinetic simulations \citep[e.g., see][]{wilsoniii21a}.  The median value of $ \omega_{\mathrm{pe}}/|\Omega_{\mathrm{e}}|$ is nearly 200. 

Unlike in the solar wind, other examples of field tangling in astrophysics, such as solar-flare turbulence, cannot be measured in situ, and so must be inferred from remote-sensing observations. In solar physics, we often use spectral-line observations from instrumentation like the Hinode EUV Imaging Spectrometer \citep[EIS;][]{2007SoPh..243...19C}, Interface Region Imaging Spectrometer \citep[IRIS;][]{2014SoPh..289.2733D}, and upcoming missions such as the MUlti-slit Solar Explorer \citep[MUSE;][]{DePontieu2022} to infer field tangling and its impact on electron transport.

This article reviews a selection of transport phenomena relating to electrons in tangled magnetic fields. It focuses on collisionless space and astrophysical plasmas, in which collective behaviour dominates the interactions between electrons and variations in the magnetic field across scales. We first introduce contemporary ideas for the creation of tangled magnetic fields. Magnetohydrodynamic (MHD) turbulence and fluid instabilities tangle magnetic fields and thus naturally create inhomogeneities in the field (Sect.~\ref{sec:Turbulence and Fluid Instabilities}). On smaller scales, kinetic instabilities tangle the magnetic field (Sect.~\ref{sec:tangled field creation by kinetic instabilities}). In this context, we also briefly point at the magnetogenesis problem and the turbulent dynamo, which relate to the creation of tangled magnetic fields from an unmagnetised initial state. Through the tension force and the exchange of magnetic stresses, however, magnetic fields resist being tangled (Sect.~\ref{Ferraro}). We briefly discuss the modulation of field-parallel heat flux by kinetic instabilities in Sect.~\ref{sect_inst_effect}. 
A key aspect of electron transport in tangled fields relates to trapping effects in inhomogeneous magnetic-field configurations (Sect.~\ref{sect:transport}), for which planetary magnetospheres provide many important example cases. We also discuss de-trapping effects due to the breaking of adiabaticity of the electron trajectories (Sect.~\ref{sect:detrapping}). Electron diffusion and energisation across the magnetic field are the scope of Sect.~\ref{sect:across}, which discusses the phenomenology of electron scattering and then presents an overview over diffusion in energy and space. The article concludes with a brief summary and outlook in Sect.~\ref{sect:conclusions}.

\section{The creation of tangled magnetic fields} \label{sec:The creation of tangled fields}

In this section, we discuss the creation of tangled fields by two broad classes of processes: those associated with fluid instabilities and turbulence (Sect.~\ref{sec:Turbulence and Fluid Instabilities}), and those generated by the conversion of energy from particles to fields through kinetic processes that disrupt particle transport (Sect.~\ref{sec:tangled field creation by kinetic instabilities}).

\subsection{Field tangling by turbulence and fluid instabilities} \label{sec:Turbulence and Fluid Instabilities}

Turbulence is present in most space and astrophysical plasmas: the solar corona \citep{Cranmer2019}, the solar wind \citep{bruno13}, the interstellar medium \citep{Armstrong1995,ferriere2020plasma,fraternale2021waves}, and the intracluster medium \citep{Fabian2006, vazza2012turbulence, cho2022effects}. In this section, we discuss the turbulent cascade, its multi-scale structure, various types of anisotropies of turbulent fluctuations, and the role of fluid instabilities.

\begin{figure}[ht]
\centering
\includegraphics[width=\textwidth]{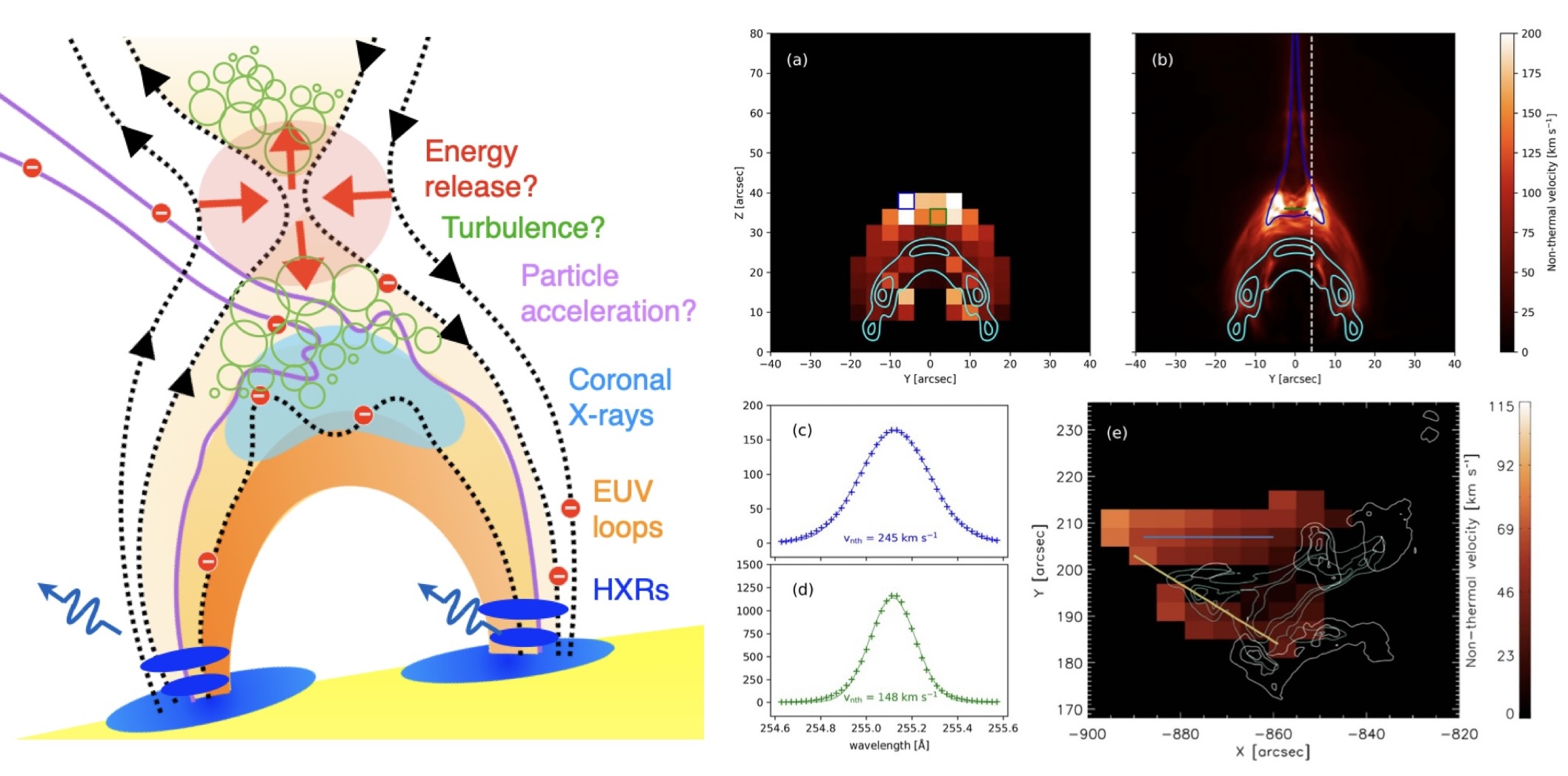}  
\caption{Left: Simple cartoon depicting the standard flare model showing possible sites of energy release, generation of turbulence in the corona, tangled fields, particle acceleration, heating, and subsequent X-ray emissions. Right: Flare observation of extended regions of turbulence in the solar corona (panel e), inferred from the presence of excess spectral-line broadening  \citep{2023ApJ...946...53S}. The observation is compared with extended regions of turbulence generated in MHD simulations of the KHI (panels a through d). Figure taken from \citet{2023ApJ...947...67R}}
\label{sf_fig1}
\end{figure}

Solar flares are one example plasma process in which field tangling by turbulence plays a crucial role, especially for the transport of electrons. Solar flares change dynamically in space and time during their evolution, possibly favouring stochastic acceleration processes \citep[e.g.,][]{1993ApJ...418..912L,2012SSRv..173..535P,2022ApJ...924...52R}, generated by turbulence and plasma waves.  Irrespective of the exact acceleration mechanism(s), turbulence is likely to play a key role. For instance, shocks require repeated particle crossings for efficient acceleration, possibly being fed by turbulence. Fluid instabilities, such as the tearing-mode instability with the generation of magnetic islands and the Kelvin--Helmholtz instability (KHI) created in the plasma sheet above coronal loop tops can generate turbulence. Moreover, ample evidence for turbulence comes from spectral-line observations, specifically broadening, and possibly from spectral-line shapes \citep[e.g.,][]{2017ApJ...836...35J}. Spectral lines are often wider than expected from the underlying ion thermal motions alone, leading to \textit{non-thermal line broadening} \citep{2015SoPh..290.3399M}. This excess broadening is often attributed to random macroscopic plasma motions and turbulence. Observations show that turbulence may play a role throughout the flare from coronal reconnection sites to the lower chromosphere. Spectral observations \citep{2014ApJ...788...26D,2021ApJ...923...40S} show distinct line-broadening patterns, such as non-thermal line broadening increasing with temperature and height in a coronal loop, or a decrease in line broadening as we move from the loop top down the loop legs (see Fig.~\ref{sf_fig1}). These observations suggest that, while turbulence may be most crucial in the loop tops, it has extended presence in multiple locations during a flare and  plays a key role in energy transfer throughout the event. In addition, spectral-line broadening occurs also within the transition region and chromosphere, even during the flare rise phase, giving a hint at the presence of turbulence  \citep{2011ApJ...740...70M,2018SciA....4.2794J}.  Other studies have shown non-thermal broadening in active regions before the flare \citep{2013ApJ...774..122H} and also at later times in long-duration flares \citep{2019ApJ...887L..34F}.
Magnetic energy must transfer from large to small scales in flares, and turbulence is an important mechanism for such a transfer of energy (see Sect.~\ref{turb_cascade}). A substantial fraction of the released energy goes into the flare-acceleration of electrons with estimates of 10-50\% \citep{2012ApJ...759...71E}. Multi-wavelength observations allow us to estimate the partitioning of  energy in flares \citep{2017PhRvL.118o5101K}. In this study, X-ray observations of the spectral-line broadening suggest that the instantaneous kinetic energy associated with turbulence is sufficient to power the acceleration of non-thermal electrons. Moreover, the estimation of the timescale for dissipation from turbulent kinetic energy to electron thermal energy is of order 1-10 seconds, similar to predictions based on  MHD modelling \citep{2017PhRvL.118o5101K}.

\subsubsection{The turbulent cascade and multi-scale structure}
\label{turb_cascade}

A flow is turbulent when the nonlinear terms in the dynamical equations describing the system -- for example, the Navier--Stokes equation in hydrodynamics, the  MHD equations for collisional magnetofluids,  or the Vlasov--Maxwell set of equations for collisionless plasmas -- are dominant and lead to a transfer of energy across scales. When the timescale associated with these nonlinear terms is comparable to or shorter than the timescales associated with linear processes, the system develops into a stochastic state \citep{Sreenivasan1992,goldreich1995}. A state of homogeneous and statistically stationary turbulence emerges when the correlation length $\ell_{\mathrm c}$ and the correlation time $\tau_{\mathrm c}$ of the auto-correlation functions of the characteristic quantities of the system are small compared to the system size and its global evolution time \citep{matthaeus1982measurement}. The nonlinear terms in the dynamic equations for these turbulent systems transfer energy across scales, which ultimately leads to dissipation at small scales \citep{bruno13, matthaeus2015intermittency, marino2023scaling}. In the classic picture of fluid turbulence \citep{kolmogorov1941}, this transfer begins at the driving scale and ends at the scales associated with collisional dissipation. The driving scale, on which the system is unstable, is comparable to $\ell_{\mathrm c}$ and thus a large scale compared to the dissipation scales. The fluctuations then become nonlinear and develop a turbulent cascade that proceeds towards smaller scales.

In magnetofluids, which evolve according to the MHD equations, the magnetic field $\vec B$ is coupled to the flow through the magnetofluid's bulk velocity $\vec U$ according to the induction equation
\begin{equation}
\frac{\partial \vec B}{\partial t}=\grad\btimes\left(\vec U\btimes\vec B\right)+\eta\nabla^2\vec B,
\end{equation}
where $\eta$ is the magnetic diffusivity. The flow is affected by the magnetic field through the Lorentz force in the MHD momentum equation. Under ideal conditions (i.e., $\eta\rightarrow 0$), the field is \emph{frozen into the flow} \citep{alfven43}. Therefore, when the flow develops turbulent fluctuations, so does the magnetic field. In this way, MHD turbulence creates tangled fields across a range of spatial and temporal scales. These in turn affect the trajectories of electrons.

Once the fluctuations in the system are nonlinear and cascade, they generate and couple to fluctuations on different scales. The kinetic energy of the flow is an ideal invariant that cascades conservatively in hydrodynamic turbulence within the so-called \emph{inertial range} of scales, which is the range of scales between driving and dissipative scales. In three-dimensional incompressible MHD turbulence, the total energy, the cross-helicity, and the magnetic helicity are the three cascading invariants \citep{matthaeus1982measurement}. 

\begin{figure}[ht]
\centering
\includegraphics[width=\textwidth]{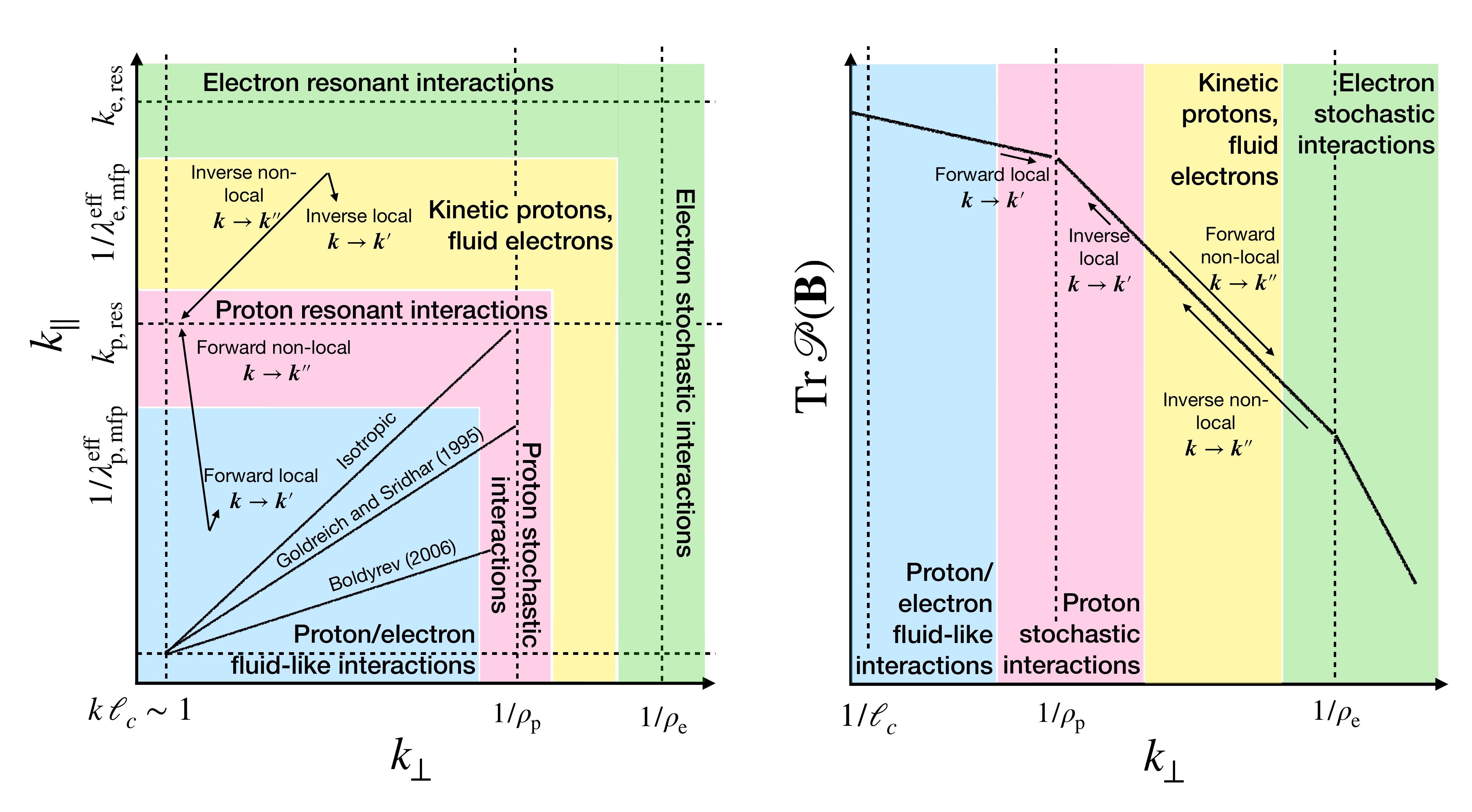}
\caption{Scale-dependent fluctuations in the magnetic field and their interactions with protons and electrons through the turbulent cascade, wave--particle resonances, and stochastic interactions. The left panel shows a two-dimensional plane of wavevector space with the field-perpendicular component $k_{\perp}$ on the horizontal and the field-parallel component $k_{\parallel}$ on the vertical axis. The non-arrowed solid lines represent theoretical predictions for the cascade of MHD turbulence. The right panel shows the trace power spectral density $\mathcal P(\vec B)$ of the magnetic field integrated over $k_{\parallel}$. The colours indicate different regimes discussed in the text. Arrows illustrate forward and inverse spectral transfer which can be local ($\vec k\rightarrow \vec k^{\prime}$) or non-local ($\vec k\rightarrow \vec k^{\prime\prime}$) in wavevector space. At scales above the effective mean free paths $\lambda_{\mathrm{p,  mfp}}^{\mathrm{eff}}$ and  $\lambda_{\mathrm{e,  mfp}}^{\mathrm{eff}}$, collisionless processes fluidise protons (blue regime) or electrons (yellow regime), respectively. Resonant interactions dominate at $k_{\parallel}\sim k_{j,\mathrm{res}}$ for species $j$ 
\label{fig:turbulence schematic}}
\end{figure}

The power spectral densities of the bulk velocity and magnetic field in MHD turbulence follow power laws in the inertial range, which is also observed in the solar wind \citep{coleman1968,bruno13}. The power spectral density as a function of wavenumber measures the distribution of the signal's power over spatial scales. For a generic signal $f(r)$, the power spectral density is $S(k)= \mathcal FR(k)$, where $\mathcal FR(k)$ is the Fourier transform of the auto-correlation function $R(\rho)=\langle f(r)f(r+\rho) \rangle_r$. The Fourier transform translates the $r$-dependence of the signal $f$ into a dependence on wavenumber $k$. This concept can be easily extended to multi-dimensional signals that depend on all three spatial coordinates $\vec r$. A schematic of a typical power spectrum is shown in Fig.~\ref{fig:turbulence schematic}. The panel on the right-hand side shows that the fluctuation amplitude decreases with decreasing scale.

Multiple models exist for the description of the cascade of MHD turbulence. In Fig.~\ref{fig:turbulence schematic}, we indicate three representative models through black solid lines: an isotropic cascade, the critical-balance model \citep{goldreich1995}, and the dynamic-alignment model \citep{boldyrev06}. The distribution of energy through the MHD cascade is a matter of ongoing research \citep{chandran2015,mallet2015,schekochihin2022}.

The large scales of the system, typically referred to as the \emph{driving scales}, provide large-amplitude fluctuations that drive the turbulent evolution of the system (shown as the scale with $k \ell_{\mathrm c} \sim 1$ in Fig.~\ref{fig:turbulence schematic}). 
In the case of the solar corona and solar wind, the reflection of Alfv\'en waves propagating away from the Sun plays a key role in the development of turbulence \citep{Heinemann1980,chandran2009alfven, bruno13, Cranmer2019, chandran2019reflection}. 
These  Alfv\'{e}n waves are launched by the footpoint motion of the magnetic-field lines in the photosphere and lower corona. When these waves propagate into gradients in the coronal background field strength and density, they undergo partial reflection that generates 
Sunward-propagating Alfv\'{e}n waves. Oppositely propagating Alfv\'{e}n waves interact nonlinearly, and the system becomes turbulent \citep{drake2013,nielson2013,pezzi2017revisiting, verniero2018}. A combination of models, simulations, and (indirect) observations provides good evidence for the wave-reflection model in the corona and near-Sun region \citep{10.1086/518001, Cranmer2019}. 
Moreover, the parametric decay of large-amplitude Alfv\'{e}n waves, in which a wave decays into two oppositely propagating Alfv\'{e}n waves and a compressive wave, leads to counter-propagating waves in the solar wind \citep{10.1086/155829, 10.3847/1538-4357/aa9bef}.  The observation that the solar wind is turbulent throughout most of the heliosphere suggests that the driving of turbulence must persist as the solar wind propagates away from the Sun \citep{Zank1996, 10.1086/498671}. At large distances from the Sun, the driving of large-scale fluctuations by interstellar pick-up ions injects fresh energy into the turbulent cascade \citep{breech08}.

At large scales, the system is in a fluid-like regime that is appropriately described through a low-order set of velocity moments of the kinetic equation \citep{verscharen2017}. Such a description is justified when the system can be approximated with a simple equation of state or when the evolution of the system does not rely on higher-order moments altogether (e.g., in non-compressive Alfv\'enic turbulence or in MHD turbulence). In collisionless plasmas, however, a low-order truncation of the velocity-moment hierarchy is not valid, and more complex kinetic models are required to capture the evolution  accurately. In collisional plasmas, this transition to the kinetic regime occurs when fluctuations have smaller scales than the collisional mean free path of species $j$ (i.e., at scales with $k\lambda_{j, \mathrm{mfp}}\gtrsim 1$). However, Coulomb collisions are not sufficient to explain the observed  fluidisation in collisionless heliospheric and astrophysical plasmas at large scales. Therefore, other processes than collisions must be responsible for the observed fluidisation of these systems at large scales. We associate these processes with \emph{effective mean free paths} in analogy to the collisional fluidisation \citep{coburn22}. We define $\lambda_{\mathrm{p,  mfp}}^{\mathrm{eff}}$ and $\lambda_{\mathrm{e, mfp}}^{\mathrm{eff}}$ for protons and electrons, respectively.
The transition from the fluid into the kinetic regime then occurs when $k_{\parallel} \lambda_{\mathrm{p,  mfp}}^{\mathrm{eff}} \gtrsim 1$ for protons and  when $k_{\parallel}\lambda_{\mathrm{e, mfp}}^{\mathrm{eff}} \gtrsim 1$ for electrons.  In Fig.~\ref{fig:turbulence schematic}, we highlight the relevant transitions and the associated regimes.

Other important scales of the system are those at which thermal particles become resonant with the fields and fluctuations of the fields perturb the particle gyro-radii, causing their trajectories to develop into a stochastic state. 
Resonant interactions require that particles and wave fields fulfil a resonance condition in which the field-parallel wavevector component $k_{\parallel}$ matches a resonance wavenumber $k_{j,\mathrm{res}}$ for particle species $j$. This resonance wavenumber depends on $v_{\parallel}$ of the resonant particles, the real part $\omega_{\mathrm r}$ of the wave frequency of the resonant waves at wavenumber $k_{\parallel}=k_{j,\mathrm{res}}$, and potentially the cyclotron frequency of the resonant particles through the resonance condition
\begin{equation}\label{rescond1}
\omega_{\mathrm r}(k_{j,\mathrm{res}})=k_{j,\mathrm{res}}v_{\parallel}+n\Omega_j,
\end{equation}
where $n$ is an integer that classifies the order of the resonance
(see also Sect.~\ref{detrapping}). 
Stochastic interactions require that the wavevector component $k_{\perp}$ perpendicular to the background magnetic field is comparable to $1/\rho_{j}$ for species $j$. Both resonant and stochastic processes can act on thermal particles or on energetic particles of species $j$ depending on the relevant $k_{j,\mathrm{res}}$ and $\rho_j$. The scales for resonant and stochastic interactions are highlighted in pink (protons) and green (electrons) in Fig.~\ref{fig:turbulence schematic}. The ordering of scales in Fig.~\ref{fig:turbulence schematic} follows the ordering for thermal particles in typical solar wind at 1\,au \citep{coburn22} but other orderings are possible in different environments or for other particle species.

Weakly collisional systems not only exhibit significant fluctuations in higher-order moments of the velocity distributions but also often develop non-equilibrium features in the background distributions of the particles that drive instabilities and thereby modify the field geometry. Unlike a turbulent cascade, these instabilities create non-local spectral transfer of fluctuation energy. In Fig.~\ref{fig:turbulence schematic}, we indicate schematically non-local direct interactions that  arise, e.g., when large-scale fluctuations cause the system to become kinetically unstable by generating temperature anisotropy \citep[cf.][]{verscharen2016collisionless,arzamasskiy23}. The instability then leads to the growth of waves at the proton gyro-scale. We discuss these processes further in Sect.~\ref{sec:tangled field creation by kinetic instabilities}. Magnetic reconnection, in turn, can lead to non-local inverse transfer of energy \citep[cf.][]{franci2017magnetic, zhou2022spontaneous, Zhou2024}, for example, through the coalescence of magnetic islands formed by reconnection.

\subsubsection{Anisotropies of the turbulent fluctuations}

The geometric structure of the tangled fields created by turbulence is not isotropic \citep{strauss1976,montgomery1981,montgomery1982,shebalin1983,higdon1984,goldreich1995,Goldreich1997, Horbury2008, bruno13}. In fact, there are multiple aspects of the turbulent fluctuations that are anisotropic with respect to the background magnetic field \citep{horbury2012}. 
The components of the fluctuating field amplitude in the direction perpendicular to the background field are typically greater than the component of the fluctuating field amplitude parallel to the background field (variance anisotropy). The energy transfer rate depends on the direction of the wavevector and is thus anisotropic (cascade anisotropy). Smaller-scale fluctuations are typically more elongated along the background field than larger-scale fluctuations, resulting in an anisotropy in the distribution of power across wavevector space (wavevector anisotropy). Lastly, the spectral index depends on the direction of the wavevector (spectral index anisotropy). The anisotropies as well as the overall power levels and polarization properties of the turbulence vary over time and are thus likely not universal \citep[see, e.g.,][]{Schekochihin2009, oughton2020critical,schekochihin2022, howes2024fundamental}. 
The anisotropy of the fluctuations has drastic implications for the field's ability to affect electron transport. Therefore, the generation of field structures and the nature of the cascade are integral parts of our understanding of tangled magnetic fields in plasmas.

\subsubsection{Fluid instabilities}

Fluid instabilities tangle the magnetic field on length scales much greater than the thermal electron scales. Therefore, electrons usually remain magnetised in these structures and follow the field lines. However, fluid instabilities can excite a turbulent cascade, bringing fluctuations to smaller scales through nonlinear interactions. In addition, energetic electrons, such as cosmic-ray electrons, may have gyro-radii of order the size of the structures created by fluid instabilities. In those cases,  large-scale structures may lead to diffusion or trapping of energetic electrons.

Fluid instabilities represent the growth of perturbations in (magneto-) hydrodynamic systems. The MHD description applies on scales associated with the driving and inertial range of the turbulence (scales in the blue shaded region of Fig.~\ref{fig:turbulence schematic}). A key instability in space and astrophysical plasmas is the KHI, which arises when there is a shear in a fluid flow \citep{helmholtz1868,kelvin1871,rayleigh1880,chandrasekhar1961}. In a magnetised fluid, the KHI naturally produces tangled fields in its nonlinear phase as it generates vortices and rolls in the fluid and the frozen-in magnetic field, ultimately leading to plasma mixing as well as mass and energy transport \citep{nykiri2001plasma,hasegawa2004transport, matsumoto2006turbulent, henri2013nonlinear, faganello2017magnetized}. An easily accessible location for the in-situ observation of the KHI is in the shear flows at the flanks of Earth's magnetopause, which is the interface between the magnetosheath and the magnetosphere \citep{hasegawa2004transport, hasegawa2006single,foullon2008evolution,taylor2012spatial,eriksson2016MMS,blasl2022multiscale}. The KHI is often observed at large (fluid) scales where it acts as an additional driver for a turbulent cascade. At the interface between Kelvin--Helmholtz vortices, ion-scale regions of intense magnetic stresses are observed \citep{sorriso2019turbulence}. On the smaller electron scales, at which the ions and electrons decouple, electron shears can lead to electron KHI \citep{fermo2012secondary, zhong2018evidence, che2023electromagnetic}. Its nonlinear phase generates tangling of field lines on electron scales and structures such as plasmoids and flux ropes.

In accretion discs, the magnetorotational instability (MRI) is a linear instability that facilitates angular-momentum transport and mass accretion in a wide range of astrophysical discs \citep{1991ApJ...376..214B,1991ApJ...376..223H,1998RvMP...70....1B}. It arises when a weak magnetic field threads a rotating disc in which the angular velocity decreases with radius, such as in accretion discs around stars or black holes. By exerting torques on magnetically tethered fluid elements, the perturbed magnetic field acts in such a way that outwardly (inwardly) displaced fluid elements gain (lose) angular momentum as they continue to move outwards (inwards). In this way, the free energy of the disc’s differential rotation is converted into radial and azimuthal motions, ultimately enabling mass accretion onto the central massive object.
More than three decades of numerical simulations have demonstrated that the nonlinear evolution of the MRI is a state of turbulence in which spatial anisotropy on large scales is structured by the differential rotation of the disc. More recent simulations reveal what appears to be an inertial-range power spectrum approaching that of Alfv\'enic guide-field turbulence at small scales \citep{walker2016,kawazura2024}.
The MRI is expected to play a critical role in many astrophysical systems, including protoplanetary discs, where it is believed to influence gas dynamics and planet formation efficiency \citep{mohanty2018,jankovic2021}. It also plays an important role in discs around compact objects like neutron stars and black holes, powering high-energy emission through efficient angular-momentum transport \citep{2015PhRvD..92f4034K}. In galactic discs, the MRI may also contribute to the generation of large-scale magnetic fields \citep{2013ApJ...764...81M}.

Heat transport itself can create tangled magnetic fields through buoyancy instabilities, provided that the transport occurs predominantly along magnetic-field lines. One of these instabilities is the magnetothermal instability  \citep[MTI;][]{2000ApJ...534..420B,balbus2001}. It arises when the plasma temperature gradient is parallel to the gravitational force (i.e., hot at the bottom, cold at the top) and leads to strongly tangled magnetic-field configurations \citep{2011MNRAS.413.1295M}. The conditions for the driving of the MTI occur in the outskirts of galaxy clusters \citep[e.g.,][]{2023A&A...680A..24K} and in hot, dilute accretion flows \citep{2011MNRAS.413.2808B}. However, the action of the MTI in weakly collisional plasmas depends critically on the plasma viscosity \citep{Kunz2011,Kunz2012} and on microphysical wave--particle interactions that affect heat transport  \citep{XuKunz2016,2024A&A...682A.125P}. 

Of similar relevance to the creation of tangled fields is the heat-flux buoyancy-driven instability \citep[HBI;][]{2008ApJ...673..758Q}. It acts when the temperature gradient is anti-parallel to the gravitational force (i.e., cold at the bottom, hot at the top), as is the case in the centres of cool-core galaxy clusters. This instability tends to re-arrange magnetic fields into a tangential configuration, suppressing field-aligned heat transport between different radii \citep{2009ApJ...703...96P, 2009ApJ...704..211B}. However, it too depends critically on the plasma viscosity \citep{Kunz2011,Kunz2012}, as well as on the amount of background turbulence driven by, e.g., feedback from the cluster's central dominant galaxy \citep{RuszkowskiOh2010}.

\subsection{Field tangling by kinetic instabilities} \label{sec:tangled field creation by kinetic instabilities}

In magnetised plasmas, kinetic instabilities create fluctuations in the magnetic field near the ion or electron gyro-radius scales, which may  affect the transport of electrons. In Fig.~\ref{fig:turbulence schematic}, the creation of these fluctuations is indicated as non-local power transfer to the scales at which resonant or stochastic interactions take place. Kinetic instabilities can modulate the magnetic-field direction and amplitude at electron (and ion) scales, lead to the creation of turbulent fields through magnetogenesis followed by a turbulent dynamo process, and create large-scale fluctuations in the magnetic field when driven by energetic particles.

\subsubsection{Ion-scale and electron-scale instabilities in magnetised plasma}\label{sect_ion_electron_inst}

Many plasma instabilities on ion or electron scales feed off pressure anisotropy as their main source of free energy. The generation of pressure anisotropy is a generic process in any magnetised, weakly collisional, and dynamic plasma system (see also Sect.~\ref{sect:magnetogenesis}).

Pressure anisotropy with respect to the background magnetic field can drive gyro-scale instabilities such as the mirror-mode \citep{1969PhFl...12.2642H, 1993JGR....98.9181S}, firehose \citep{chandrasekhar58,rosenbluth58,parker58b,vedenov61,1996JGR...10124457Q,1998JGR...10314567G}, ion-cyclotron \citep{kennel1966,1992JGR....97.8519G}, and whistler-wave instabilities \citep{kennel1966,1996JGR...10110749G}. These instabilities serve to limit  the pressure anisotropy of the particle velocity distribution functions by creating ion-scale or electron-scale fluctuations that, in their nonlinear phase, trap and/or scatter particles in pitch angle. Although these instabilities do not produce strongly tangled magnetic fields per se, the mirror-mode instability can give rise to ion-gyro-scale fluctuations with $\delta B/B \sim 0.3$ \citep{2014PhRvL.112t5003K,2015ApJ...800...27R}. These fluctuations are capable of generating magnetic holes in which electrons can be trapped, affecting the transport of electrons \citep{2016MNRAS.460..467K,2016ApJ...824..123R,liu25} and modifying the electron velocity distribution within the holes \citep{2022ApJ...935..169J,2024ApJ...965..155L}. Section~\ref{sec: trap_mirror} discusses these trapping effects in more detail.

Many ion-scale and electron-scale instabilities are also associated with collisionless shock waves, and indeed, many strong electron-scale fluctuations are commonly observed near shock waves. A collisionless plasma shock is most often a nonlinearly steepened fast-magnetosonic/whistler wave that has reached a balance between steepening and dispersion. The resultant discontinuity continually generates free energy from the driver of the shock.  Since these shocks are a steepened form of a dispersive wave mode, the shock itself can radiate waves on the same branch of the dispersion relation \citep[][]{tidman68a,krasnoselskikh02a}, known as \emph{whistler precursors}.  Whistler precursors can also be generated by shock-reflected particles through a modified two-stream instability \citep[][]{matsukiyo06a}.  These waves have wavelengths that span from above ion scales to electron scales  \citep[][]{hull12a, hull20a, wilsoniii12a, wilsoniii13a, wilsoniii14a, wilsoniii14b, wilsoniii17a}.
Whistler precursors can also greatly affect the incident flow (both ions and electrons), and within these flow modulations, even smaller-scale electrostatic waves can be generated \citep[][]{wilsoniii07a, wilsoniii14a, wilsoniii14b, wilsoniii21a}.  The same free-energy source that generates whistler precursors at strong shocks (i.e., through particle reflection) can generate much smaller-scale waves as well.  Waves radiated by electron-cyclotron drift instability (ECDI) at shocks \citep[e.g.,][]{breneman13a, wilsoniii10a, wilsoniii14b} are direct evidence of electron--ion coupling since the relative drift between reflected ions and incident electrons acts as their free-energy source.  The waves heat suprathermal electrons perpendicular to the quasi-static magnetic field, thermal electrons parallel to the field, and generate parallel ion tails through the coupling between Doppler-shifted ion-acoustic modes and electron Bernstein modes \citep{chapter6}.

\subsubsection{Magnetogenesis and the turbulent dynamo}\label{sect:magnetogenesis}

The tangling of magnetic fields is often related to two scientific challenges: (1) the origins of magnetic fields in the first place, and (2) the means by which these seed fields are amplified to dynamical strengths. The interplay between kinetic instabilities and turbulence forms a possible scenario to explain this chain of events that leads to tangled magnetic fields.

When considering the first challenge, the Weibel instability \citep{1959PhRvL...2...83W} is a promising candidate mechanism to provide a seed magnetic field. This instability is driven by anisotropic particle pressure configurations in (nearly) unmagnetised plasmas. Small magnetic-field fluctuations with wavevectors oriented in the direction of the colder part of the velocity distribution reinforce fluctuations in the current density oriented in the direction of the hotter part of the distribution.  Once unstable, the plasma grows fluctuations in the magnetic field until the field becomes strong enough to magnetize the plasma and thereby affect the trajectories of the electrons. Additional trigger mechanisms may be in place, for instance two-stream instabilities that create Langmuir waves due to the differential streaming of electrons and ions in regions with spatially modulated particle densities \citep{Schlickeiser2003}. Resonant wave--particle interactions relax the two-stream distributions to be anisotropic, which then triggers a secondary Weibel instability. An initially imposed velocity shear can also generate electron anisotropy, which then drives the Weibel instability \citep{Pucci2021,zhou2022spontaneous}.

After the creation of the seed field through the Weibel instability, a turbulent dynamo process sets in that leads to the further growth of the magnetic field, until the field growth is balanced with dissipation, for instance through magnetic reconnection \citep{sironi2023}. This proposed dynamo process addresses the second challenge of magnetogenesis: the growing of the seed field into a dynamically important large-scale magnetic field.

\begin{figure}[ht]
\centering
\includegraphics[width=\textwidth]{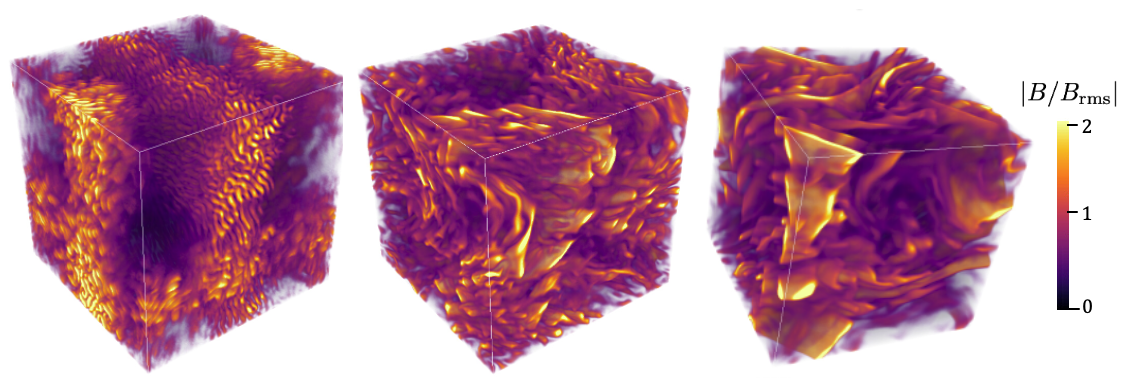}
\caption{Magnetogenesis through the Weibel instability and the turbulent dynamo.
The three panels show snapshots from a kinetic particle-in-cell simulation. The colour indicates the magnitude of the magnetic field normalised to its root-mean-square value. Left: at the time of peak growth of the Weibel instability. Middle:  after one large-scale turnover time. Right: saturated state of the dynamo. From \citet{Zhou2024}}\label{fig:Zhou_weibel}
\end{figure}

Figure~\ref{fig:Zhou_weibel} demonstrates the involved stages in a fully kinetic particle-in-cell simulation  \citep{Zhou2024}. The left panel shows the simulation domain at the time when the Weibel instability is at its peak growth, and the growing kinetic-scale fluctuations  are apparent. The field growth saturates as the fluctuations disrupt the electron gyro-motion. Next, the filamentary field coalesces, and the saturated field is transported to larger scales through magnetic reconnection. The result of this process is visible in the middle panel of Fig.~\ref{fig:Zhou_weibel}, which shows a snapshot of the simulation domain after one turbulent turnover time. The system then transitions to the dynamo saturation stage, shown in the right panel of Fig.~\ref{fig:Zhou_weibel}. 

\begin{figure}[ht]
\centering
\includegraphics[width=\textwidth]{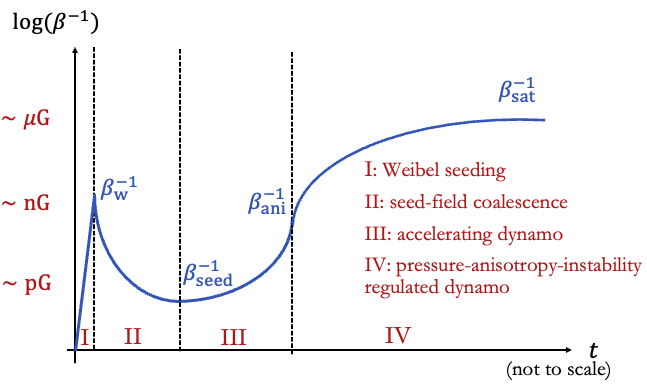}
\caption{The magnetogensis scenario proposed by \citet{Zhou2024}. The red labels on the vertical axis indicate typical field strengths in astrophysical plasmas during the evolutionary phases of this scenario, and $\beta$ is the ratio between the thermal and magnetic pressure. Phase~I describes the creation of the seed field through the Weibel instability. In Phase~II, the field undergoes coalescence and magnetic reconnection \citep[cf.][]{zhou2022spontaneous}. In Phases~III and IV, the plasma is in a collisionless dynamo state, which is ultimately regulated by pressure-anisotropy instabilities  \citep{Rincon2016,StOnge2018}. From \citet{Zhou2024}} \label{fig:Zhou_magnetogensis}
\end{figure}

Figure~\ref{fig:Zhou_magnetogensis} depicts the time evolution of the phases (from Phase~I to Phase~IV) proposed in this magnetogenesis scenario.
Future simulations must seek to connect consistently the seed and growth problems of magnetogenesis by fully realising a simulation that begins in Phase~I and ends in Phase~IV. At the end of Phase~III, the large-scale fluctuations of the magnetised plasma trigger mirror-mode and firehose instabilities, which self-regulate the pressure anisotropy in the plasma at later stages. This self-regulation acts like an effective viscosity that controls the parallel rate of strain of the plasma and the rate of magnetic-field amplification \citep{Rincon2016,StOnge2018}. 
A dynamo in a collisionless plasma with a field-strength-dependent effective collisionality can lead to the rapid growth of the seed magnetic field \citep{Schekochihin2006,Mogavero2014}.
At all stages of this proposed scenario for magnetogenesis, tangled fields occur and play a crucial role for the transport of electrons.

\subsubsection{Field generation by energetic particles}

In addition to field generation by thermal particles, also energetic particles can drive instabilities that generate and modulate magnetic fields. For instance, cosmic rays escaping from their sources along the local  magnetic field can drive a variety of streaming instabilities, including two resonant types: the gyro-resonant instability \citep{1969ApJ...156..445K,2018MNRAS.476.2779L,2019ApJ...882....3H,2019ApJ...876...60B,2021ApJ...920..141B,2021ApJ...914....3P} and the intermediate-scale instability \citep{2021ApJ...908..206S, 2023JPlPh..89f1703S, 2025ApJ...979...34L}. These instabilities influence cosmic-ray transport in the interstellar medium, the intracluster medium, and near cosmic-ray sources such as supernova remnants.

Resonant streaming instabilities result in small magnetic-field amplifications with $\delta B/B \ll 1$ in environments like the interstellar and intracluster media \citep{2019ApJ...882....3H}. However, near cosmic-ray sources such as  upstream of non-relativistic shocks, magnetic amplification by resonant streaming instabilities can reach $\delta B/B \sim 1$ due to the high energy density of the streaming cosmic rays in these environments, as shown by kinetic shock simulations \citep{2014ApJ...794...46C}.

Upstream of shocks with sufficiently high Mach numbers, a different branch of the streaming instability, the non-resonant streaming instability, dominates \citep{2004MNRAS.353..550B,2009MNRAS.392.1591A,matthews17}. In this regime, the magnetic-field amplification reaches $\delta B/B \gg 1$ as shown in simulations \citep{2014ApJ...794...46C}. In the presence of the large energy density of the streaming cosmic rays, the instability grows on scales much smaller than the cosmic-ray gyro-radius, and thus the amplified field has a minimal impact on the particle trajectories. Consequently, the electric current of the cosmic rays, which drives the instability, remains weakly affected by the growing fluctuations, allowing for continued growth and thus eventually large field amplifications \citep[e.g.,][]{2009ApJ...694..626R}. 

As the growing modes become nonlinear, the cosmic-ray gyro-radius decreases due to the increasing field strength while the non-resonant instability transitions to larger scales at which large-scale instabilities may further contribute to tangle the magnetic field.  This change eventually affects the cosmic-ray trajectories, producing saturation when the magnetic power is concentrated at scales near the cosmic-ray gyro-radius.

The non-resonant streaming instability occurs for any orientation of the cosmic-ray current with respect to the background magnetic field \citep{2005MNRAS.358..181B}. In particular, the case in which the cosmic-ray current is perpendicular to the background field has the potential to generate highly tangled fields \citep{2010ApJ...717.1054R}. Due to its large field amplification, the non-resonant streaming instability can significantly reduce the diffusion coefficients of both energetic ions and electrons in the upstream medium of non-relativistic shocks, impacting the maximum energy of shock-accelerated particles \citep{2012MNRAS.419.2433R,2014ApJ...794...47C} as well as their spectra \citep{2020ApJ...905....2C,2021A&A...650A..62C,2021ApJ...922....1D}.

Cosmic-ray pressure effects may cause the inflation of large bubble-like structures near cosmic-ray sources as potential sites of high-intensity gamma-ray emission \citep{schroer2021dynamical,schroer2022cosmicray}.

\subsection{The magnetic field's resistance to being tangled}\label{Ferraro}

Magnetic fields, once generated, resist being tangled.  In MHD, the magnetic tension force ${\propto}(\vec B\bcdot\grad)\vec B$ accelerates the conducting fluid, and thus the frozen-in magnetic-field lines, so as to straighten the field.
One consequence of this effect for rotating, axisymmetric, ideal MHD systems is captured by \emph{Ferraro's isorotation theorem}. It states that the angular velocity of a conducting fluid must be constant along magnetic-field lines in steady state \citep{ferraro37}. As an application of the theorem, magnetospheres with poloidal magnetic fields around planets and stars co-rotate with their central bodies when in equilibrium. 
A poloidal field cannot exist in equilibrium without co-rotation because the frozen-in plasma would wind up the magnetic field in this case and build up magnetic tension. It would create toroidal field out of the initial poloidal dipole field by advection \citep{ogilvie16}.

In the case of a planetary magnetosphere, the ionosphere co-rotates with the planet due to collisional friction with the atmosphere. As per Ferraro's theorem, the co-rotating plasma of the ionosphere would stretch the magnetic field and create a time-dependent state of the field in the magnetosphere. Instead, if the Alfv\'en-wave propagation timescale is smaller than the dynamical timescale, electric currents effectively communicate  magnetic stresses along the planetary magnetic-field lines, and isorotation is more likely to hold. In this case, the field remains poloidal and the plasma remains co-rotating. 
The same effect occurs in pulsar magnetospheres, although here relativistic effects impact the communication between the footpoints of the field on the neutron star and the magnetosphere
\citep{goldreich69,uzdensky03}.

Ferraro's isorotation theorem and these examples illustrate how magnetic fields resit tangling in plasma systems through tension forces and the exchange of magnetic stresses.
In the case of the solar wind, the interplanetary magnetic field is not poloidal and winds up by advection \citep{parker58}. It develops toroidal components and magnetic tension that exchange angular momentum between the field and the flow, ultimately slowing down the rotation of the Sun \citep{weber67,verscharen21}.

\section{Modulation of parallel electron heat flux by kinetic instabilities}\label{sect_inst_effect}

In plasmas with low levels of collisions, electrons often carry significant heat flux due to their high mobility \citep{hollweg74,feldman76}. The electron heat-flux vector is defined as the third moment of the electron velocity distribution function $f_{\mathrm e}$:
\begin{equation}
\vec q_{\mathrm e}=\frac{m_{\mathrm e}}{2}\int \left(\vec v-\vec U_{\mathrm e}\right)\left(\vec v-\vec U_{\mathrm e}\right)^2f_{\mathrm e}\,\mathrm d^3v,
\end{equation}
where $\vec U_{\mathrm e}$ is the bulk velocity (i.e., the first velocity moment) of $f_{\mathrm e}$. Electron heat transfer is dominated by heat flow along magnetic field lines. When plasma processes create a non-zero divergence of $\vec q_{\mathrm e}$, the heat flux contributes to  a local change in thermal energy density \citep{chapter7}. Therefore, the behaviour of heat flux is of great interest to our understanding of plasma energetics. 

In the solar wind, the electron heat flux can be large \citep{scime99,cranmer21,halekas21} and  drive plasma instabilities \citep{gary75,gary99,verscharen22}. These collisionless processes compete with the collisional regulation of the heat flux according to the Spitzer--H\"arm theory \citep{spitzer53} in the solar wind \citep{bale13}.
In a wide range of astrophysical plasmas, heat conduction plays an important role, for instance, during galaxy formation and in the intracluster medium \citep[e.g.,][]{2002MNRAS.335L..71F}.  
In the case of the intracluster medium, thermal conduction from the hot outskirts into the radiatively cooling core may help stave off a cooling catastrophe \citep{2003ApJ...582..162Z,2003ApJ...596L.139K}, which would otherwise lead to catastrophic mass accretion rates of up to a thousand solar masses per year \citep{1977MNRAS.180..479F}. In fact, X-ray spectroscopic observations of galaxy clusters \citep{2002MNRAS.335L..71F} indicate typical cold-mass accretion rates onto the central dominant galaxy of tens, and at most hundreds, of solar masses per year \citep[e.g.,][]{2003ApJ...590..207P}, a puzzle known as the \emph{cooling-flow problem} \citep{pf06}.

Complicating the thermal regulation of the cooling intracluster medium are indications that heat conduction in such environments can be suppressed by factors of $\sim0.1$ to $\sim10^{-2}$ relative to the Spitzer--H\"arm prediction \citep{bc81,2000MNRAS.317L..57E, 2003ApJ...586L..19M}. Similar levels of heat-flux suppression are also observed in laboratory experiments \citep{2022SciA....8J6799M} and first-principles numerical simulations (see \citealt{chapter7} for a review). All these transport effects occur in tangled and turbulent magnetic-field configurations, which further complicate any reductions in the potency of the conductive heat flux. 
Understanding whether and to what extent electromagnetic fields are ordered or tangled in different space and astrophysical contexts therefore has deep implications for heat transport and for the acceleration of, for instance, cosmic ray electrons (and ions) in shock waves \citep[e.g.,][]{morlino2021particle} and by second-order (stochastic) Fermi processes \citep{brunetti2020secondorder} in various heliospheric and astrophysical environments.

All instabilities discussed in Sect.~\ref{sect_ion_electron_inst} as well as an additional family of electron-resonant instabilities \citep{verscharen22} have the potential to modify electron transport parallel to the magnetic field and thus also the electron heat flux. A detailed discussion of electron-mediated heat transport in space, astrophysical, and laboratory plasmas by collisions and collisionless processes including  kinetic instabilities is provided in the companion review to this article by \citet{chapter7}. Electron-driven instabilities and their impact on the evolution of the electron distribution function are discussed in the companion review to this article by \citet{chapter4}.

\section{Electron trapping in inhomogeneous magnetic fields}\label{sect:transport}


We now describe the dynamics of electrons that are trapped in a local minimum of the magnetic-field strength while the gyro-centres of their trajectories move parallel or anti-parallel to the field.
An electron passing adiabatically through such an inhomogeneous magnetic field changes its pitch-angle $\theta$, where $\cos\theta = v_{\parallel}/v$. While the electron conserves its kinetic energy $W$ and magnetic moment $\mu=W\sin^2\theta/B$, the modulation of $\theta$ is determined by the spatially varying magnetic-field strength $B$. In a local depletion of $B$, a magnetic bottle can form, which traps part of the electron distribution.

The modulation of electron trajectories due to a changing magnetic field causes changes in the electron velocity distribution. The evolving velocity distribution can, especially before achieving a steady state, contain sufficient free energy to excite microinstabilities.
In this case, trapping is an example of a process through which structures with large scales (ion scales or larger) couple with electron-scale structures through the kinetics of the electron distribution.
Similar physical processes occur over a range of scales and in different plasma environments. In this section, we show examples from the Earth's  magnetosphere and magnetosheath as these environments provide directly observable space plasmas in which electron trapping plays an important role. 

\begin{figure}[ht]
    \centering
    \includegraphics[width=\textwidth]{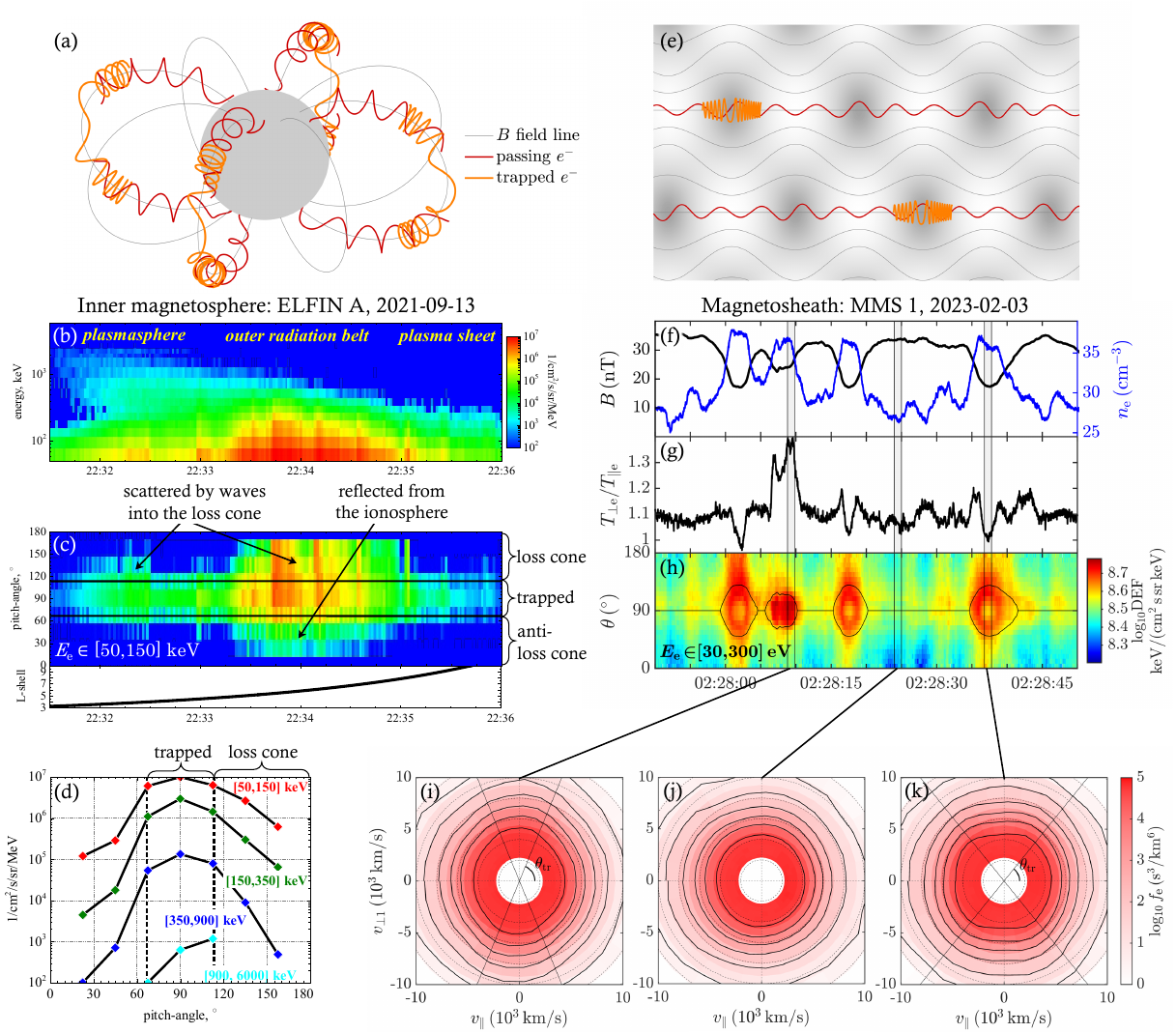}
    \caption{Trapping in inhomogeneous magnetic-field structures. (a) Passing/escaping (red) and trapped (orange) electrons in a planetary dipole magnetic field.
    (b) Spectra of trapped electrons measured by the ELFIN CubeSat \citep{Angelopoulos20:elfin} at different locations \citep[for details, see][]{Artemyev22:jgr:ELFIN&THEMIS}.
    (c) Pitch-angle distribution of $[50,150]$\,keV electrons for the same interval as in (b) with indication of trapped, scattered, and reflected populations \citep[for details, see][]{Mourenas21:jgr:ELFIN}.
    (d) Pitch-angle distribution measured by ELFIN around 22:33:41~UT.
    (e) Trapping in ion-scale magnetic structures. The gray shading indicates $B$, the gray curves represent magnetic-field lines. The electron distribution is separated into passing/escaping (red) and trapped (orange) electrons.
    (f) MMS measurements of an ion-scale structure in the magnetosheath with anti-correlation in  $B$ and $n_{\mathrm e}$, typical for mirror modes.
    (g) Electron temperature anisotropy $T_{\perp\mathrm e}/T_{\parallel\mathrm e}$ for the same interval as in (f).
    (h) Differential energy flux of $[30,300]$\,eV electrons for the same interval as in (f).
    (i)--(k) Electron velocity distribution functions $f_{\mathrm e}(v_\parallel,v_{\perp1})$ at the three times indicated in (h)
    }
    \label{fig:trapping}
\end{figure}

\subsection{Trapping in the inner magnetosphere}

The Earth’s inner magnetosphere and the magnetospheres of planets with strong magnetic fields in general are natural large-scale magnetic traps due to the inhomogeneity of their magnetic fields.
Electrons gyrate around the approximately dipolar field lines of the magnetosphere. The magnetic field strength $B$ is at a minimum in the equatorial plane and at maximum at the foot points of the field lines near the planet's poles. 

In a strong magnetic field, the electron gyro-radius $\rho_{\mathrm e}$ is sufficiently small to neglect the gyro-motion in the description of the particle dynamics. Effects due to the finite $\rho_{\mathrm e}$ in the context of magnetic trapping are discussed in Sect.~\ref{sec:current_sheets}.  Under the small-$\rho_{\mathrm e}$ assumption, we discuss two types of motion: particle oscillations along magnetic-field lines and particle cross-field drift due to curvature and gradient effects \citep{bookNorthrop63}. Due to the azimuthal symmetry of magnetic dipole field, cross-field drift results in a quasi-periodic particle motion around the planet (for a discussion of exceptions, see Sect.~\ref{sec:trapping_in_magnetotail}). We thus focus on the bounce motion in the context of particle trapping and de-trapping. 

Figure~\ref{fig:trapping}a shows examples of trapped and untrapped particle orbits in the inner magnetosphere.
In the absence of spatial and temporal perturbations on scales comparable to the gyro-radius, the magnetic moment $\mu$ is conserved \citep{bookLL:mech,Lichtenberg&Lieberman83:book}. The particle bounce motion is then a one-dimensional oscillation in an effective potential well $\mu B(s)$, where $B(s)$ is the magnetic field magnitude as a function of the field-aligned spatial coordinate $s$. We define $s=0$ as the location at which the magnetic field assumes its minimum value; i.e., at the magnetic equator. 

This configuration allows us to use an analogy with the classical nonlinear pendulum \citep{Lichtenberg&Lieberman83:book}. We introduce two regions in phase space, depending on the energy $W$ and pitch-angle $\theta$ of the particles. We define a maximum $B$-value $B_{\max}$ in such a way that particles with energy $W$ greater than $\mu B_{ \max}$ escape from the oscillatory motion. These escaping particles are illustrated by the red trajectories denoted as \emph{passing} in Fig.~\ref{fig:trapping}a. 
Particles with energy $W<\mu B_{\max}$  remain in their oscillatory motion and are thus trapped between the two locations $s_{\pm}$, which are defined by $W=\mu B(s_\pm)$. The trapped particles are illustrated by the orange trajectories in Fig.~\ref{fig:trapping}a.
For planetary magnetospheres, $B_{\max}$ is generally given by the magnetic field magnitude at those altitudes at which particle collisions with the dense ionosphere and neutrals result in particle losses (e.g., about $\sim 100$\,km in the Earth’s magnetosphere). 

Taking into account that, in a dipole field, $B$ reaches its minimum value at the equator,  the passing/escaping population consists of particles with small equatorial pitch-angles $\theta < \theta_{\mathrm{tr}}$ and $\theta > (180^\circ - \theta_{\mathrm{tr}})$, where 
\begin{equation}\label{trapping_ang}
\sin\theta_{\mathrm{tr}}(s)=\sqrt{\frac{B(s)}{B_{\max}}}
\end{equation}
defines the trapping angle $\theta_{\mathrm{tr}}(s)$
The trapped electrons are those with larger pitch-angles ($\theta_{\mathrm{tr}} < \theta < 180^\circ - \theta_{\mathrm{tr}}$) that are reflected when $B$ increases towards its maximum value $B_{\max}$.

As the passing electrons are ultimately lost to the collisional ionosphere and atmosphere, the result of this process is an electron distribution that consists only of the trapped population at all $|s|<|s_{\pm}|$. The empty part of phase space with smaller pitch-angles defines the \emph{loss cone}. The pitch-angle $\theta_{\mathrm{tr}}$ of the boundary of the loss cone at a given $s$ depends on the magnetic field strength $B_{\max}$ at the mirror point $s_{\pm}$ and the magnetic field strength $B(s)$ at position $s$ according to Eq.~(\ref{trapping_ang}); i.e., the angle depends on latitude and drift shell in a planetary field \citep[e.g.,][]{koskinen2022}.

\begin{figure}[ht]
    \centering
    \includegraphics[width=0.5\textwidth]{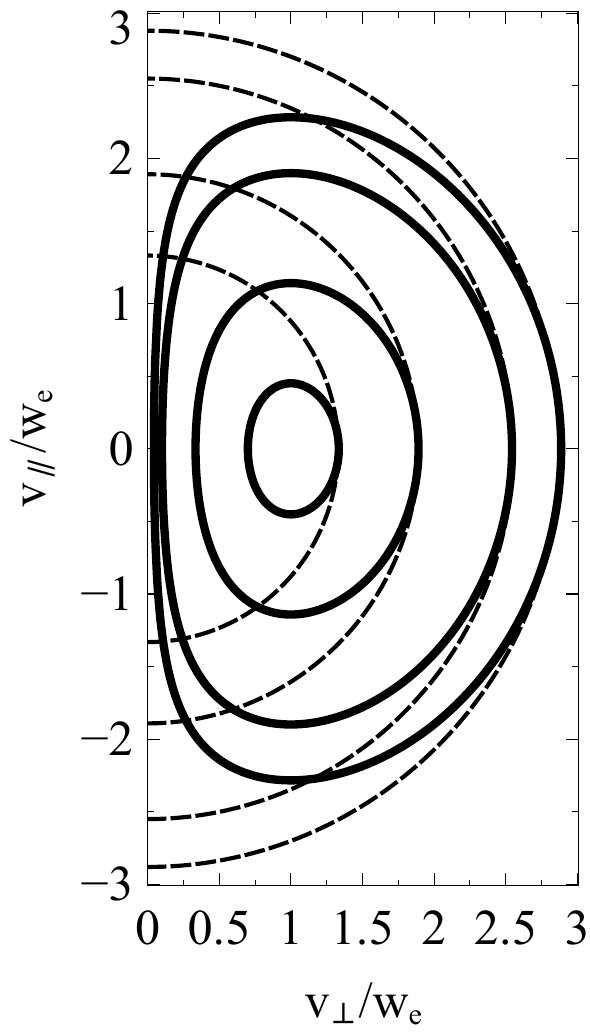}
    \caption{Illustration of the loss-cone distribution in Eq.~(\ref{losscone}) with $a=1$ (solid curves). For comparison, we overplot the corresponding Maxwellian distribution without a loss cone ($a=0$, dashed curves). The distributions are visualised as contours of constant phase space density }
    \label{fig:LC}
\end{figure}

A simple analytical form of the loss-cone distribution is given by
\begin{equation}
\label{losscone}
f_{\mathrm e}\left( v_{\perp},v_{\parallel}\right) = \frac{{n_0 v_ \perp ^{2a} }}{{\pi ^{3/2} a\Gamma (a)w_{\mathrm e}^{3 + 2a} }}\exp \left( { - \frac{{v_{\perp}^2+v_\parallel ^2 }}{{w_{\mathrm e}^2 }}} \right),
\end{equation}
where $\Gamma(x)$ is the $\Gamma$-function and $a>0$ is a constant parameter that determines the size of the loss cone (see Fig.~\ref{fig:LC}). 

The bounce period is given by \citep{bookLyons&Williams}
\begin{equation}\label{taub}
\tau_{\mathrm b}=\sqrt{2m_{\mathrm e}}\int\limits_{s_-}^{s_+}\frac{1}{\sqrt{W-\mu B(s)}}\,\mathrm ds=\sqrt{\frac{2m_{\mathrm e}}{W}}\int\limits_{s_-}^{s_+}\frac{1}{\sqrt{1-\displaystyle\frac{B(s)}{B(0)}\sin^2\theta }}\,\mathrm ds,
\end{equation}
where the second equality follows from $\mu=W\sin^2\theta/B(0)$. For relativistic particles, $2m_{\mathrm e}/W\to 1/c^2$ in Eq.~(\ref{taub}), and then the bounce period depends  on pitch-angle only but not on energy. We find that $\tau_{\mathrm b}\sim\ell/\sqrt{2W/m_{\mathrm e}}$ where $\ell\approx (s_+-s_-)$ is the system's spatial scale along magnetic-field lines. Figure~\ref{fig:trapping} shows measurements of  trapped and escaping populations of electrons  at different distances from the equator in the Earth’s magnetosphere.

\subsection{Trapping in ion-scale and electron-scale magnetic-field structures}
\label{sec: trap_mirror}

The concept of a loss cone can be generalised to any magnetic-field configuration with a local minimum in $B$. Thus, loss-cone-like distributions arise also in other local magnetic-field structures, such as mirror modes or magnetic holes. Figure~\ref{fig:trapping}e illustrates the trapping and passing of electrons in a quasi-period ion-scale structure (e.g., a chain of mirror modes). The magnetic-field strength $B$ follows a spatial modulation, represented by the varying levels of gray shading. Due to the scale separation between the size of the ion-scale structure and $\rho_{\mathrm e}$, the electrons follow adiabatic trajectories in this field configuration. Electrons with $\theta_{\mathrm{tr}}<\theta<(180^{\circ}-\theta_{\mathrm{tr}})$ are trapped in the ion-scale structures, while electrons with $\theta< \theta_{\mathrm{tr}}$ or $\theta>(180^{\circ}- \theta_{\mathrm{tr}})$ pass through the ion-scale structures.

Figure~\ref{fig:trapping}f shows an example of mirror modes measured by the MMS spacecraft in Earth's  magnetosheath behind a quasi-perpendicular bow shock. Mirror modes are non-propagating pressure-balanced structures consisting of compressive magnetic-field fluctuations (peaks and troughs in the field magnitude) accompanied by anti-correlated fluctuations in thermal pressure (seen as density troughs and peaks).
The characteristic size of mirror modes is of order the ion gyro-radius and typically observed on timescales of 3-24\,s in the magnetosheath \citep{soucek2008}. In the associated magnetic-field troughs, part of the electron population are trapped and form a loss-cone-like distribution. 

Figure~\ref{fig:trapping}h shows the modulation of the electron pitch-angle distribution in the  mirror-mode structures shown in Fig.~\ref{fig:trapping}f. The black curves represent the trapping angle as $\theta_{\mathrm{tr}}$ and $(180^\circ-\theta_{\mathrm{tr}})$ from Eq.~(\ref{trapping_ang}). The part of the distribution between these curves corresponds to trapped particles, while the part outside these curves corresponds to the loss cone of passing particles. We take $B_{\max}$ to be the average $B$ across the interval \citep[see][]{yao2018}. Below panel (h), we  show two-dimensional cuts in $(v_\parallel,v_{\perp1})$ space of the electron velocity distribution $f_{\mathrm e}$ at three times during the interval, where $v_{\perp1}$ is one of the Cartesian velocity axes in the plane perpendicular to $
\vec B$. The straight lines in these figures indicate the separation between trapped (at small $|v_{\parallel}|$) and passing (at large $|v_{\parallel}|$) particles. The first distribution (Fig.~\ref{fig:trapping}i) shows that the  trapped population in the magnetic minimum of this mirror mode exhibits larger values of $f_{\mathrm e}$ at large pitch-angles ($\theta_{\mathrm tr}<\theta<180^\circ-\theta_{\mathrm{tr}}$). This behaviour translates to a high electron temperature anisotropy with $T_{\perp\mathrm e}/T_{\parallel\mathrm e}>1$ (see also Fig.~\ref{fig:trapping}g).

The second distribution (Fig.~\ref{fig:trapping}j) is taken outside the mirror mode. It is more isotropic than inside the field depletion as shown by the solid black isocontours that lie close to the dotted black contours of constant energy. Consequently, this distribution has a lower $T_{\perp\mathrm e}/T_{\parallel\mathrm e}$ than the distribution inside the structure.

\begin{figure}[ht]
    \centering
    \includegraphics[width=0.7\textwidth]{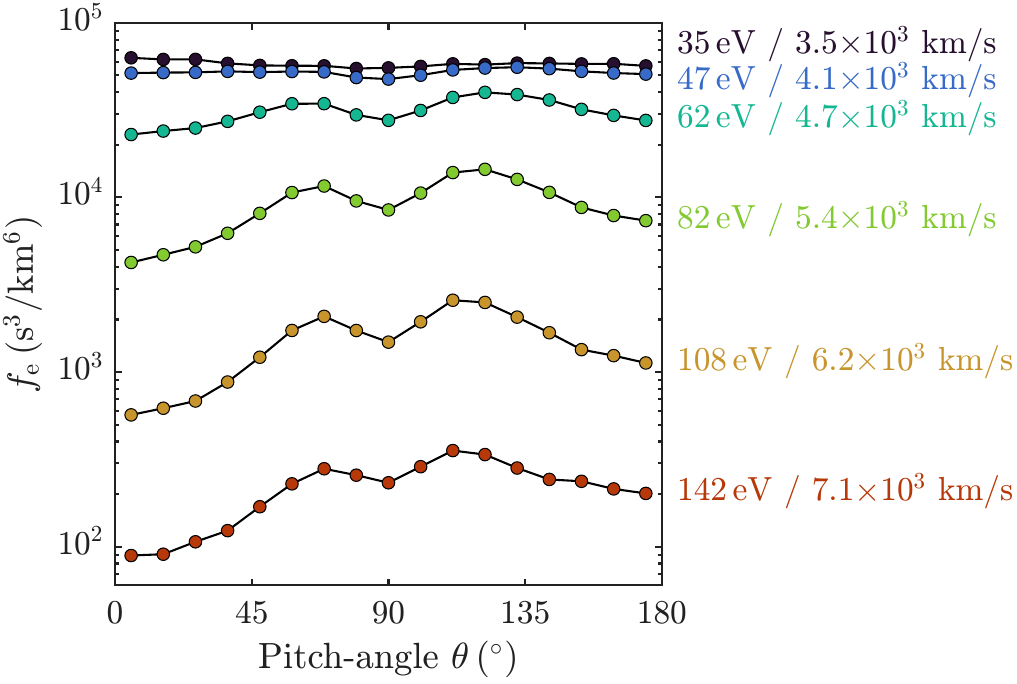}
    \caption{Butterfly distribution measured by MMS in the Earth's magnetosheath, same distribution as in Fig.~\ref{fig:trapping}. Each curve shows $f_{\mathrm e}$ as a function of pitch-angle $\theta$ for different electron energies/speeds as indicated by colours.
     }
    \label{fig_butterfly}
\end{figure}

Magnetic minima can also reshape the electron distribution in other ways. An example is shown by the third distribution (Fig.~\ref{fig:trapping}k). Here, $f_{\mathrm e}$ is reduced at small pitch-angles, $\theta < \theta_{\mathrm tr}$ and $\theta >(180^{\circ}- \theta_{\mathrm tr})$, like in the loss cone shown in the first distribution. The contours of $f_{\mathrm e}$ also exhibit a constriction  at $\theta \approx 90^\circ$. This distribution corresponds to the so-called \emph{butterfly} or \emph{doughnut distribution}, which is often found in mirror-mode structures in the magnetosheath downstream the quasi-perpendicular bow shock \citep{yao2018} and in local magnetic-field minima downstream the quasi-parallel bow shock \citep[e.g.,][]{svenningsson2022}. Figure~\ref{fig_butterfly} shows an example of such a butterfly distribution for different energies. A possible explanation for this shape involves the combination of betatron and Fermi acceleration and deceleration processes as the mirror mode grows deeper \citep{yao2018,2022ApJ...935..169J}.
Despite the presence of the loss cone and other strong non-equilibrium features in the distribution in Fig.~\ref{fig:trapping}k, $T_{\perp\mathrm e}/T_{\parallel\mathrm e} \approx 1$. 

It is important to consider the exact shape of the distribution when studying secondary instabilities, as two distributions with the same $T_{\perp\mathrm e}/T_{\parallel \mathrm e}$ can indeed have very different phase-space features and thus kinetic properties including their potential to drive microinstabilities.
For instance, the trapping of electrons can excite  whistler waves within mirror-mode structures, which in turn re-shape the electron distribution \citep{kitamura2020,2022ApJ...935..169J,svenningsson2022,jiang2024}. 
In the solar wind, when a core--strahl electron configuration encounters ion-scale magnetic holes, the suprathermal part of the electron distribution can be modulated in such a way that it emits Langmuir waves through the bump-on-tail instability \citep{boldu23,liu25}. Through the scattering with the growing Langmuir waves, some electrons then transition from the passing strahl population into the trapped population. 
These examples show that self-regulated wave--particle interactions through secondary instabilities can play a critical role in the multi-scale dynamics of trapped electrons in nonlinear structures.

Butterfly distributions are also regularly observed in Earth's radiation belts; however, it is likely that other processes are responsible for the minima at pitch-angles near $90^{\circ}$ in these regions. At large $L$-shells, they are probably caused by drift-shell splitting and magnetopause shadowing \citep{ni20}. At low $L$-shells, they are caused by local acceleration due to wave--particle interactions \citep{xiao2015}.

Electron trapping can also occur in coherent electron-scale structures that are frequently observed in space plasmas \citep{huang17,yao17,liu19,zhou19} and sometimes occur nested inside ion-scale structures \citep{li20}. For instance, observations by MMS in the Earth's magnetosheath reveal flux enhancements in the electron pitch-angle distribution near the local maximum of the magnetic-field strength \citep{xie24}. Trapping by the magnetic mirror force cannot explain this behaviour. Instead, an inhomogeneity of the electric potential along the direction of the magnetic field associated with these structures may potentially explain this electrostatic trapping effect at electron-scale structures. In this scenario, the electric potential in combination with the mirror force transforms the initially isotropic electron distribution into a trapped and a streaming population.

\subsection{Trapping by transient magnetic-field enhancements} \label{sec:trapping_in_magnetotail}

A global dipole magnetic field causes charged particles to follow a closed azimuthal drift motion around the planet along contours of constant equatorial field \citep[see][]{bookRoederer70}. In open systems, however, gradient drifts can result in particle losses through the system boundaries.
The Earth's bow shock and the Earth's magnetotail are two examples of such boundaries. Here, the spatial limitations of the systems along the drift direction can result in particle escape. 

An effective mechanism of particle trapping in such systems occurs in the form of local magnetic-field peaks, changing the drift trajectories and thus trapping particles in analogy to closed drift trajectories in dipole fields. Simulations reveal this kind of trapping of ions and electrons in the Earth’s magnetotail  \citep{Gabrielse17,Ukhorskiy18:DF}. Nonlinear dipolarisation fronts associated with transient magnetic reconnection form the magnetic field enhancements in this case  \citep[see][]{Nakamura04, Runov09grl,Sitnov09}. 

Similar trapping mechanisms may also manifest in  planetary foreshocks when spatially localised transients form magnetic-field enhancements \citep[e.g.,][]{Omidi10, Turner13:foreshock}. The main feature of this trapping process is the transport of the trapped particles along with the magnetic-field enhancements. The particles can be energised through adiabatic betatron heating if the transport carries them across increases in the background magnetic field \citep{Gabrielse17,Ukhorskiy18:DF}. 

Wave--particle interactions can lead to the de-trapping of particles from such magnetic-field configurations. When undergoing these resonant interactions, particles do not conserve their magnetic moment any more, leading to a change in their drift orbit or to  scattering due to magnetic field curvature (see Sect.~\ref{sec:current_sheets}).

\subsection{De-trapping of electrons}\label{sect:detrapping}

Particles that are trapped in steady-state field structures remain trapped as long as their motion is adiabatic. However, wave--particle interactions, microinstabilities, the betatron effect, scattering by magnetic-field rotations, and collisions can transfer particles from the trapped population into the escaping loss cone of the distribution. These processes lead to the de-trapping of particles.

\subsubsection{De-trapping through wave--particle interactions}\label{detrapping}

Particles can transition from the trapped to the escaping region of velocity space through particle scattering with fluctuating electromagnetic fields. In order for these fields to disrupt the conservation of the magnetic moment $\mu$, they often have spatial and/or temporal scales comparable to the scales of the gyro-motion of the particles.  Resonant wave--particle interactions are a sub-family of such processes \citep{bookSchulz&anzerotti74,bookLyons&Williams}.

Electrons resonate with a wave when they fulfil the resonance condition \citep{Lichtenberg&Lieberman83:book,bookLyons&Williams}
\begin{equation}\label{rescond}
\omega_{\mathrm r}=k_{\parallel}v_{\parallel}+n\Omega_{\mathrm e}
\end{equation}
introduced more generally in Eq.~(\ref{rescond1}).
For $n=0$, the resonance condition describes the Landau resonance; for all other $n$, it describes cyclotron resonances of different orders.
Equation~(\ref{rescond}) suggests that waves with frequency $\omega_{\mathrm r} \sim n|\Omega_{\mathrm e}|$ are very effective in scattering particles, viz., by covering a large range in resonant $v_{\parallel}$. Indeed, in the Earth's inner magnetosphere, electrons are efficiently scattered by electron-harmonic  waves with $\omega_{\mathrm r}\sim n|\Omega_{\mathrm e}|$ and $n=1,2,3,\dots$ \citep[e.g.,][]{Zhang15:ECH,Ni16:ssr}, whistler waves with $\omega_{\mathrm r}\lesssim|\Omega_{\mathrm e}| $ \citep[e.g.,][]{Mourenas12:JGR,Ni16:ssr},  fast-magnetosonic/whistler waves with $\omega_{\mathrm r} \ll |\Omega_{\mathrm e}|$ and $\omega_{\mathrm r}\sim k_\parallel v_\parallel$  \citep{Horne07,Mourenas13:JGR:magnetosonic}, and electrostatic broad-band waves with $\omega_{\mathrm r} \sim n|\Omega_{\mathrm e}| $  and $n=1,2,3,\dots$ \citep{Vasko18:pop,Shen20:jgr:tds}. 

\begin{figure}[ht]
    \centering
    \includegraphics[width=\textwidth]{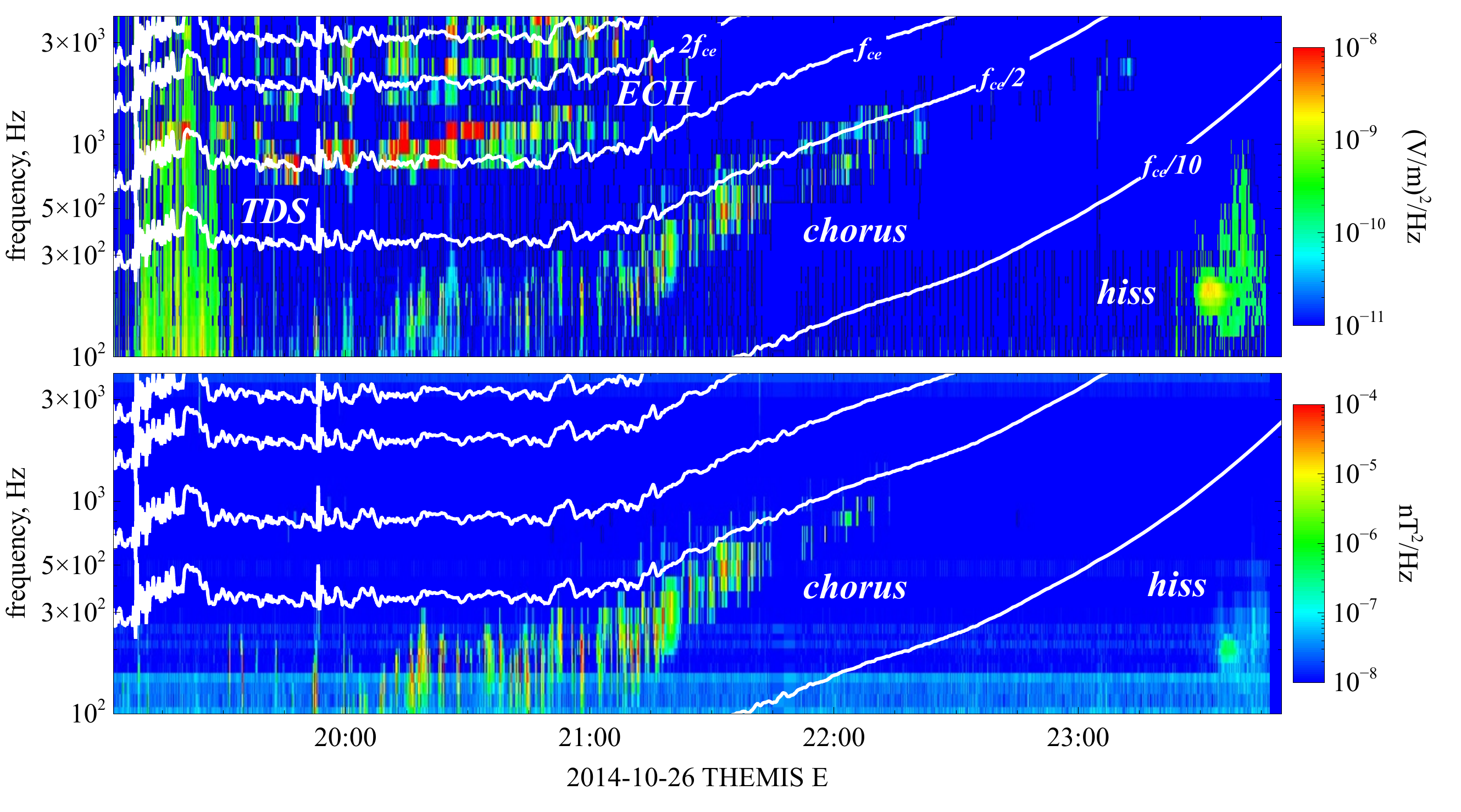}
    \caption{Spectrograms of the main wave modes responsible for electron scattering into the loss cone in the inner magnetosphere measured by THEMIS~E \citep{Angelopoulos08:ssr}. The top panel shows the electric-field spectrum, and the bottom panel shows the magnetic-field spectrum \citep{Bonnell08,LeContel08}. The white curves show electron-cyclotron harmonics $nf_{\mathrm{ce}}$ and fractions $f_{\mathrm{ce}}/n$, where $f_{\mathrm {ce}}= |\Omega_{\mathrm e}|/2\pi$, measured by the fluxgate magnetometer \citep{Auster08:THEMIS}. The main wave modes are identified, see text for definitions
     }
    \label{fig:waves}
\end{figure}

Figure~\ref{fig:waves} shows a measured spectrogram of wave modes that affect the electron dynamics and result in the scattering of electrons into the loss cone.\footnote{In general, $\Omega_{\mathrm e}\propto 1/\gamma$, where $\gamma=1/\sqrt{1-v^2/c^2}$, so that electrons at high (relativistic) energies  have a sufficiently small $|\Omega_{\mathrm e}|$ to resonate with low-frequency waves such as electromagnetic ion-cyclotron waves \citep{Kersten14,Ni15}.} 
Time-domain-structures \citep[TDS;][]{Mozer15} are electrostatic nonlinear solitary waves observed as broad-band noise that effectively scatters electrons with $W<10$\,keV  \citep{shen2024}. Part of this broad-band electrostatic noise can be due to short-wavelength kinetic Alfv\'en waves \citep{Chaston12} that effectively scatter  electrons with $W\sim 100$\,keV \citep{Shen23:jgr:ELFIN_THEMIS}. Electron-cyclotron harmonics \citep[ECH;][]{Zhang15:ECH} are electrostatic high-frequency waves that effectively scatter  electrons with $W<1$\,keV \citep{Ni16:ssr}. Whistler chorus and hiss waves \citep{Agapitov13:jgr} effectively scatter electrons with energies between 10 and 1000\,keV \citep{Artemyev13:angeo}.
Similar resonant wave--particle interactions are responsible for ion scattering in the Earth's inner magnetosphere. Ion-cyclotron waves with $\omega_{\mathrm r}\leq \Omega_{\mathrm i}$ and fast-magnetosonic/whistler waves with $\omega_{\mathrm r}\sim n\Omega_{\mathrm i}$, where $\Omega_{\mathrm i}$ is the cyclotron frequency of the ion species under consideration,  effectively scatter ions \citep{Ma19:emic&ions} and transfer them into the loss cone. Pitch-angle scattering and resulting de-trapping of electrons by whistler waves can also occur in the transition layer of Earth's bow shock \citep{katou19,amano20,lindberg24}. 

The diffusion coefficient
\begin{equation}
D_{\theta\theta} (\theta) = \frac{\langle (\Delta\theta)^2\rangle}{\tau}
\end{equation}
measures the efficiency of resonant particle scattering in pitch-angle by plasma waves \citep{Andronov&Trakhtengerts64,Kennel&Engelmann66}. When evaluated at the trapping angle, $D_{\theta\theta} (\theta_{\mathrm{tr}})$ quantifies the corresponding efficiency of particle losses from the magnetic trap. In this definition, $\Delta\theta$ is the pitch-angle change for a single resonant interaction, and $\tau$ is the average time between two successive resonant interactions. $D_{\theta\theta}$ is measured in degrees$^2$/second or radians$^2$/second. The timescale of particle de-trapping, i.e., the time it takes diffusion to transport particles into the loss cone, $\theta\to\theta_{\mathrm{tr}}$, is approximately $\tau_{\mathrm L}\sim1/ D_{\theta\theta} (\theta_{\mathrm{tr}})$ when $D_{\theta\theta}$ is measured in radians$^2$/second \citep{Albert&Shprits09:JASTP}.  

Quasilinear theory is a framework that allows us to derive diffusion coefficients under the assumption that each $\Delta\theta$ is much smaller than $\theta$ and that the particle orbits are small perturbations to the unperturbed particle trajectories in the background field $\vec B_0$ \citep{Drummond&Pines62,Vedenov62}. The resulting diffusion coefficient is linearly proportional to the wave intensity, $D_{\theta\theta} \propto B_{\mathrm w}^2$, where $B_{\mathrm w}$ is the magnetic-field amplitude of the resonant wave \citep{bookSchulz&anzerotti74,bookLyons&Williams}. 
The applicability of the quasilinear approach is largely constrained to small wave intensities, $B_{\mathrm w}^2/B_0^2\ll 1$, so that the magnetic-field perturbations (and the electric-field perturbations) remain much smaller than the background magnetic field $B_0$. For a sufficiently long time $\sim \tau_{\mathrm L}$, even these small perturbations transfer a significant number of particles from the magnetic trap into the loss cone. The approximation of unperturbed trajectories is violated in the case of very intense waves \citep[for thresholds of wave amplitudes, see][]{karpman1974,Zhang22:natcom}, when nonlinear resonant effects become important \citep{Albert13:AGU,Artemyev24:ssr:arXiv}.

\begin{figure}[ht]
    \centering
    \includegraphics[width=\textwidth]{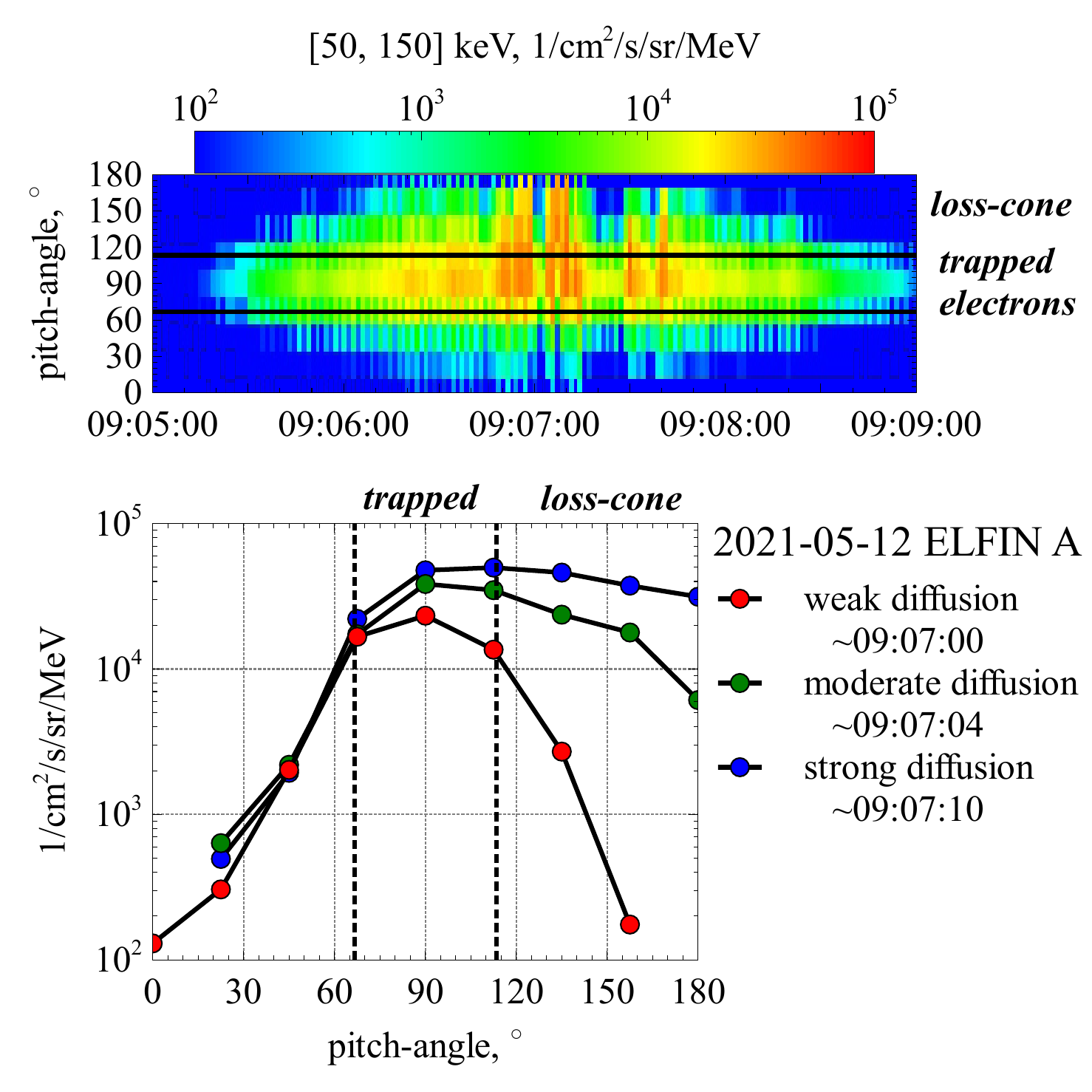}
    \caption{Typical electron pitch-angle distributions in the inner magnetosphere measured by ELFIN  \citep{Angelopoulos23:ssr}. The top panel shows pitch-angle distributions around afternoon magnetic local time. Bursts of electron fluxes within the loss cone are due to electron scattering by whistler waves \citep{Tsai22}. The bottom panel shows three pitch-angle distributions with different filling rates of the loss cone. The distribution with the full loss cone is likely formed by electron scattering in the strong diffusion limit}
    \label{fig:precipitations}
\end{figure}

The strongest rate of particle losses from  magnetic traps occurs in the \emph{strong diffusion regime}, i.e., when $\tau_{\mathrm L}$ is comparable to the bounce period $\tau_{\mathrm b}$. This regime is characterized by $D_{\theta\theta}\approx \theta_{\mathrm{tr}}^2/\tau_{\mathrm{b}}$ \citep{Kennel69} and $\theta_{\mathrm{tr}}={\rm arcsin}\left(B(0)/B_{\max}\right) \approx B(0)/B_{\max}\ll 1$. In this regime, the loss cone with $\theta<\theta_{\mathrm{tr}}$ and $\theta>(180^{\circ}-\theta_{\mathrm{tr}})$ is continuously filled with particles through a flow of particles from the magnetic trap. This is the strongest loss regime that can be described by diffusion, and it is achieved when $B_{\mathrm w}$ is sufficiently large. A further increase of the wave intensity would not change the flow of particles from the trap. Figure~\ref{fig:precipitations} shows electron pitch-angle distributions in regimes of weak and strong diffusion for electrons scattered by whistler waves in the Earth's inner magnetosphere. For very large wave intensity, the quasilinear diffusion approximation breaks down, and other nonlinear resonant effects overfill  the loss cone through a flow of particles from the trap that exceeds the strong diffusion limit \citep{Zhang22:natcom}.

\subsubsection{Self-limitation of trapping}

An important concept related to magnetic traps and particle escape from these traps is the self-limitation of trapping. The magnetic-trap configuration assumes the presence of a loss cone  that is empty or depleted in the absence of strong particle scattering. Therefore, the particle velocity distribution is not isotropic and contains a flux gap within a small pitch-angle range. This non-equilibrium feature leads to a loss-cone anisotropy, which is a source of free energy for instabilities \citep{Sagdeev&Shafranov61}.
Trapped electrons then self-consistently generate the waves responsible for their de-trapping by scattering as described in Sect.~\ref{detrapping}. This self-induced scattering into the loss cone occurs, for example, when the temperature anisotropy is large \citep{kennel1966} or when butterfly distributions are present \citep{kitamura2020,svenningsson2022}. The bottom panels of Fig.~\ref{fig:trapping} show example cases of non-equilibrium distributions in a trapping field configuration. Since these types of distributions can result from trapping, whistler waves generated by instabilities are regularly found in mirror-mode structures \citep{smith1976}. Estimates of pitch-angle diffusion coefficients reveal that whistler waves significantly impact the electron velocity distributions throughout the Earth's magnetosheath \citep{svenningsson2024}.

Under many plasma conditions, the loss-cone distribution in Eq.~(\ref{fig:LC}) is unstable to wave generation. The free energy for these instabilities is comparable to the free energy due to the associated temperature anisotropy \citep{Shklyar09:review}. Magnetic-trap configurations with such a loss-cone electron distribution are often unstable to whistler-wave and electron-cyclotron harmonic wave instabilities \citep{kennel1966,Karpman75:ECH,Maha&Kennel78,Liu18:ech}, which self-consistently scatter particles into the loss cone. For a fixed loss-cone size determined by the magnetic-field configuration, the wave growth rate is proportional to the density of the trapped particles, and thus stronger waves are generated when there are more particles inside the trap. This self-regulating mechanism  prevents the flux of trapped particles to increase unboundedly \citep{kennel1966,Summers09,Mourenas24:jgr:ELFIN&KPlimit}.

The betatron effect provides an additional self-limiting process for trapped particle fluxes in  magnetic traps. Any adiabatic variation of the magnetic configuration of the magnetic trap, i.e., a slow change of the $B(s)$ profile in time, results in adiabatic variations of the charged particle energy. While conserving the magnetic moment $\mu$, the betatron effect modifies the energy of the trapped electrons, mediated by an electric field that must obey Faraday's law,
\begin{equation}
\vec{\nabla}\times\vec E=-\frac{1}{c}\frac{\partial \vec B}{\partial t}
\end{equation}
associated with the slow time-dependence of $\vec B$. This electric field does work on the trapped particles and modifies their energy $W$.
The conservation of the second adiabatic invariant $\propto \oint{\sqrt{W-\mu B}\,\mathrm ds}$, which is associated with the bounce motion \citep{bookSchulz&anzerotti74},
results in a variation of the parallel particle energy, analogously to Fermi acceleration \citep{Bogachev05,Artemyev12:jgr:electrons,Borissov16,Lichko&Egedal20}. Such an adiabatic energy change results in the formation of anisotropic distribution functions that can become unstable to wave generation \citep{2022ApJ...935..169J,Frantsuzov24:apj}. The subsequent particle scattering by the unstable waves also leads to losses from the magnetic trap as described in Sect.~\ref{detrapping}.

In most situations, particle distributions in trapping field configurations have higher particle fluxes outside the loss cone, so that  wave--particle interactions mostly result in particle scattering towards and into the loss cone. However, resonant particle scattering can also work in the opposite direction, transferring particles from the loss cone into the trapped regime. In this case, the wave--particle interactions support and enhance the trapped population \citep{liu25}, and even allow particles to spend more time trapped in systems with large loss cones. This process is important for  particle acceleration at shock waves. Particles crossing the shock or reflecting at the shock are lost and do not return for further acceleration. However, if these particles are scattered by waves after their interaction with the shock, they may return, providing them with an opportunity for further acceleration \citep{Amano22:review}.

\subsubsection{Electron scattering by magnetic-field rotations}\label{sec:current_sheets}

\begin{figure}[ht]
    \centering
    \includegraphics[width=\linewidth]{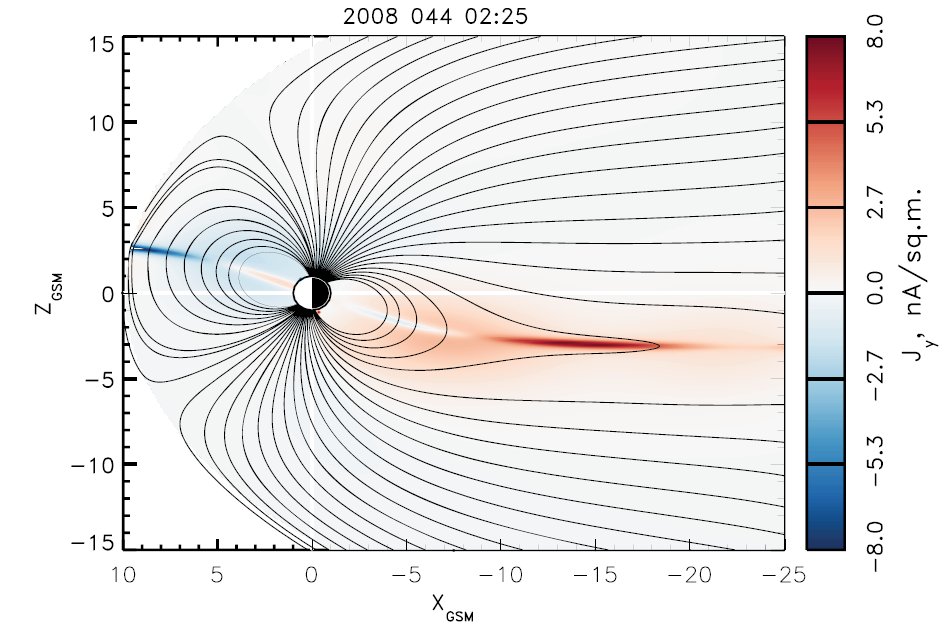}
    \caption{Magnetic-field configuration and current density in Earth's magnetosphere. The magnetic field lines are shown in black and the out-of-plane current density in colour. The most intense current density occurs in the equatorial magnetotail, where magnetic-field lines have small curvature radii. Particle scattering at small magnetic-field curvature radii de-trap particles in the magnetosphere.  Adopted from \citet{Sitnov19} and with permission by John Wiley and Sons}
    \label{fig:cs}
\end{figure}

Besides the scattering of particles by electromagnetic waves, magnetic field curvature also has the potential to scatter particles \citep{bookChirikov87}. This scattering is known as \emph{current-sheet scattering} since large rotations in the magnetic field are associated with strong currents according to Amp\`ere's law. In the magnetosphere, a large-scale current sheet where this effect is prevalent is associated with the magnetic-field curvature in the equatorial magnetotail, see Fig.~\ref{fig:cs}.

The efficiency of the current-sheet scattering mechanism for electrons is controlled by the ratio of the curvature radius $R_{\mathrm c}$ of the magnetic-field lines to the electron gyro-radius $\rho_{\mathrm e}$. If $R_{\mathrm c}/\rho_{\mathrm e}\gg 1$, the scattering is exponentially weak with $D_{\theta\theta} \propto \exp(-R_{\mathrm c}/\rho_{\mathrm e})\ll 1$ \citep{Birmingham84,Neishtadt00}. If $R_{\mathrm c}/\rho_{\mathrm e} \ll 1$, the scattering is weak and depends on $R_{\mathrm c}/\rho_{\mathrm e}$ as $D_{\theta\theta} \propto R_{\mathrm c}/\rho_{\mathrm e} \ll 1$ \citep{BZ89}. If $R_{\mathrm c}/\rho_{\mathrm e}\sim 1$, the scattering is most effective, providing an almost stochastic motion of the particles \citep{Horton97, Zelenyi13:UFN}.
When acting, this scattering mechanism traps or de-traps particles. Several examples of current-sheet scattering are found in planetary magnetotails \citep[e.g.,][]{Delcourt94:scattering,Sergeev12,Artemyev24} and in the solar wind \citep[e.g.,][]{Artemyev20,Malara21}.

\subsubsection{Electron scattering by collisions}\label{sec:collisions}

If the rate of particle collisions between the trapped particles and other particles (ions, electrons, and/or neutral particles) is sufficiently high, collisions are capable of de-trapping particles by scattering them into the loss cone. In this case, elastic and inelastic scattering changes the pitch-angle and energy of the trapped particles, eventually resulting in particle escape from the trap. 
Energetic electrons in the Earth's inner radiation belt, for example, are trapped very close to the planet: at less than 3 Earth radii in the equatorial plane. These electrons encounter dense ionospheric plasma consisting of heavy ions and free electrons during their bounce motion \citep[see][and references therein]{Selesnick12, Selesnick16}. The resulting collisions between the energetic electrons and the ionospheric plasma lead to the de-trapping of some of the energetic radiation-belt electrons.

\begin{figure}[ht]
    \centering
    \includegraphics[width=\textwidth]{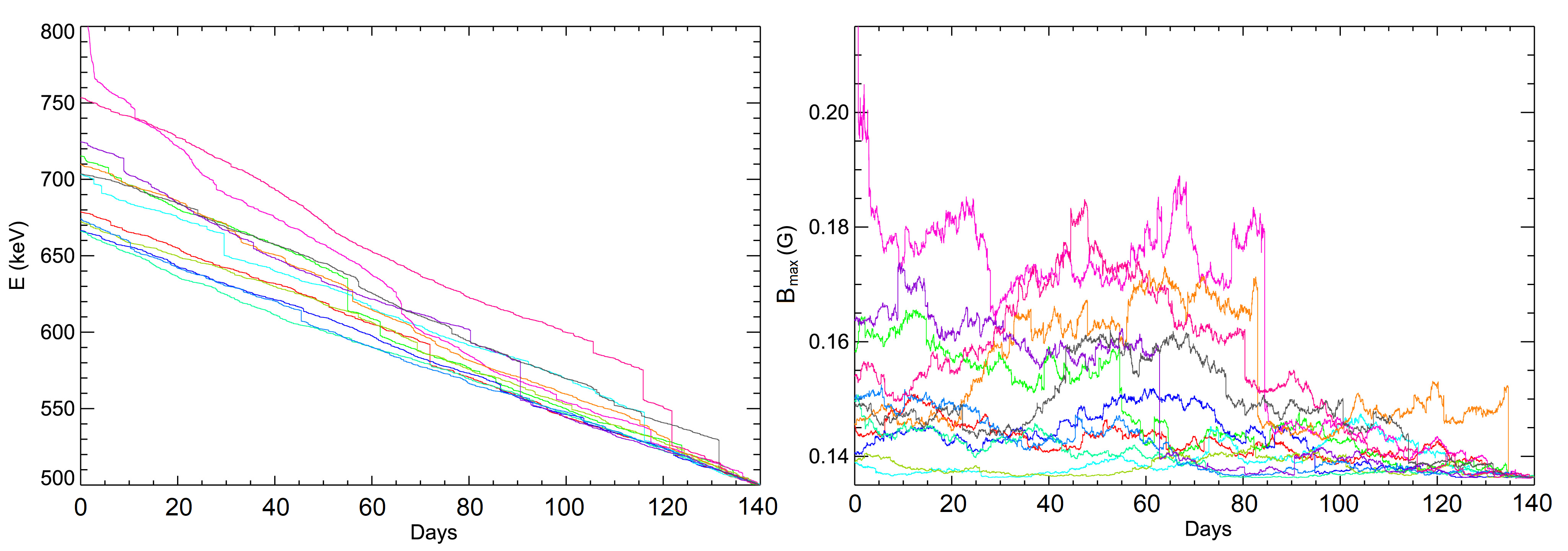}
    \caption{Results of Monte-Carlo simulations of electron trajectories undergoing elastic and inelastic collisions. Left: electron energy; right: magnetic-field strength at the particle mirror points. Electrons are trapped within the inner radiation belts and bounce between the equator with $B(0)\approx 1.39$\,G and $B_{\max}$. Adopted from \citet{Selesnick16} and with permission by John Wiley and Sons}
    \label{fig:collisions}
\end{figure}

Elastic electron collisions with ions do not change the energies $W$ of the participating electrons but scatter their  pitch-angles $\theta$. The differential cross section is the key parameter controlling the efficiency of this pitch-angle scattering. It describes the probability of a scattering event at a specific solid angle and depends on the electron energy $W$, scattering angle $\Delta \theta$ (larger for smaller scattering angle), and some properties of the participating ions or neutral particles \citep{FernandezVarea93}. 
Inelastic scattering reduces the electron energy $W$ by an amount that is used to excite or ionise the collision partners. The corresponding cross section of this interaction depends on the electron energy loss $\Delta W$ \citep[see][and references therein]{Selesnick16}.  

The combination of elastic and inelastic collisions results in decreases in the energies and random variations in the pitch-angles of the affected electrons. The change in pitch-angle can be expressed in terms of the change of the electron mirror-point position, which is quantified in terms of their associated $B_{\max}=B(0)/\sin^2\theta$. If collisional scattering results in increases in $B_{\max}$ to values associated with very low altitudes, the corresponding electrons reach the dense ionosphere at typical altitudes of $\sim 100$\,km, where they are finally lost due to the de-trapping through multiple collisions. Figure~\ref{fig:collisions} shows the results of Monte-Carlo simulations of energetic electrons trapped in the inner radiation belt of the Earth’s magnetopshere. The particle energies slowly decrease due to inelastic collisional scattering. When $B_{\max}$ sufficiently increases, particles are lost to the ionosphere. 
Collisional de-trapping also operates in the solar corona \citep[e.g.,][]{Emslie78, Kontar15} and in the radiation belts of other planets \citep[e.g.,][]{Clark14, Nenon18}.

\section{Electron diffusion and energisation across the magnetic field}
\label{sect:across}

In this section, we focus our discussion on energy diffusion and spatial diffusion of electrons across tangled magnetic fields.

\subsection{Phenomenology of electron scattering}

There are multiple ways for electrons to interact with fluctuations in the electromagnetic field. Effective \emph{non-resonant scattering} of a given electron on fluctuations in the magnetic field occurs when $k_{\perp}\rho_{\mathrm e}\sim 1$, where $k_{\perp}$ is the perpendicular component of the wavevector of the involved field fluctuations \citep{chen2001,white2002}. 
If $k_{\perp}\rho_{\mathrm e}\ll 1$, the electron follows adiabatic behaviour in these structures.
If $k_{\perp}\rho_{\mathrm e}\gg 1$, the small-scale modes average out over one gyro-period and thus do not significantly affect the electron trajectory. Figure~\ref{fig:turbulence schematic} depicts these three regimes in the context of turbulent magnetic and electric fields.

In the case of non-resonant scattering, classical Gaussian diffusion models predict that the mean squared displacement of the affected particles is proportional to the elapsed time, i.e., 
\begin{equation}
\langle\Delta x^2\rangle\propto \tau,
\end{equation}
like in the case of Brownian motion, where $\Delta x=x(t+\tau)-x(t)$. 
A complication arises from the fact that the turbulent fluctuations on which the particles scatter are intermittent and anisotropic (see Sect.~\ref{turb_cascade}). Intermittency and anisotropy of the turbulence lead to anomalous diffusion with 
\begin{equation}
\langle\Delta x^2\rangle\propto \tau^{\alpha},
\end{equation}
where $\alpha\ne1$ \citep{zimbardo2006,perri2008}. Intermittency also causes large jumps in energy space (Levy flights) and thus anomalous energy diffusion \citep{isliker2017}. Sections~\ref{sect:energy} and \ref{sect:spatial} provide further details on energy diffusion and spatial diffusion.

The transport of electrons is further complicated by resonant and stochastic behaviour. We have introduced resonant wave--particle interactions in the context of particle de-trapping in Sect.~\ref{detrapping}. If the fluctuations that participate in these wave--particle interactions propagate (i.e., $\omega_{\mathrm r}\ne0$), they can match the resonance condition with electrons in Eq.~(\ref{rescond1}) 
leading to efficient energy transfer \citep{Kennel&Engelmann66,rowlands66,Lichtenberg&Lieberman83:book}. If the fluctuations have sufficient amplitudes, they perturb the gyro-motion  significantly and scatter the particles stochastically \citep{brunetti2007compressible}.
Ion-acoustic waves often satisfy $k \lambda_{\mathrm{De}}\sim 1$, which corresponds to wavelengths that are $\sim20-100$ times less than $\rho_{\mathrm e}$ \citep[e.g.,][]{fuselier84a,breneman13a,wilsoniii21a,vasko22a}. Since these ion-acoustic waves have high frequencies, their phase speed is often comparable to the electron thermal speed, so that a large fraction of thermal electrons can go into Landau resonance with these waves despite their small wavelengths \citep[e.g.,][]{dum74a, petkaki06a}.  Likewise, Langmuir waves and lower-hybrid waves often occur with wavelengths much less than $\rho_{\mathrm e}$ and still affect electrons significantly via Landau resonance \citep[e.g.,][]{cairns05a, wilsoniii12a,yoon16b}.

\subsection{Energy diffusion}\label{sect:energy}

Plasma turbulence creates tangled electromagnetic fields (see Sect.~\ref{sec:Turbulence and Fluid Instabilities}) and is one of the key mechanisms responsible for efficient particle energisation \citep{fermi1949origin, fermi1954galactic}.  Significant numerical efforts are being devoted to the investigation of the interplay between these different mechanisms \citep{zank2015particle, zank2021flux, nakanotani2021interaction,trotta20233D} in the energisation of particles.

Turbulent fluctuations perform second-order stochastic particle acceleration in the form of a diffusion process in velocity space. In this process, some particles gain energy while others lose energy during their interaction with the fluctuating electric field. Since a larger fraction of particles increase their energy rather than decrease their energy, the process induces a net average particle energisation. 
Stochastic acceleration of relativistic electrons occurs in the intracluster medium and in galaxy clusters \citep{brunetti2001particle, petrosian2007nonthermal, brunetti2016stochastic, brunetti2020secondorder}. Observations of diffuse radio emission, an indirect proxy for energised electrons, suggest that electron energisation takes place both in the turbulent high-$\beta$ plasma of the intracluster medium and in galaxy clusters, where shock waves can induce particle energisation \citep{stevens1999galaxies, vanweeren2017case}. In the intracluster medium, the resultant heating helps to overcome the radiative cooling that affects electrons more than protons as they suffer from radiative losses \citep{Field1965}. Recently, the observation of \emph{radio bridges} connecting two galaxy clusters is interpreted in terms of stochastic acceleration due to turbulence \citep{brunetti2020secondorder}, although
other mechanisms, such as weak shocks, may contribute as well to particle energisation in the intracluster medium \citep{vanweeren2017case, ryu2019diffusive}.

In-situ observations of electron beams near the terrestrial bow shock and interplanetary shocks suggest that stochastic electron acceleration at plasma shock waves is strongly affected by ripples in the shock surfaces \citep{lindberg23,jebaraj23}. Ripples in the shock front enable multiple encounters of the electron beam with the strong gradients of the shock surface \citep{xu15}, a process to which also the herringbone structures in radio bursts are attributed \citep{holman83,zlobec93}. A complementary scenario invokes the interaction of a planar shock front with tangled and meandering magnetic-field lines that are convected over the shock surface \citep{decker93,jokipii07,guo10,guo15}. The electrons follow these tangled magnetic field lines throughout the heliosphere \citep{moradi19,bian21,li23} and thus also on the upstream (and downstream) sides of shock surfaces. Like in the shock-ripple scenario, this geometry allows for multiple encounters of streaming electrons with the shock and hence iterative acceleration. Given the ubiquity of turbulence and tangled magnetic fields in collisionless plasmas as well as the known sub-structure of shock waves, the interaction of electrons with shock surfaces is likely to be defined by a combination of both scenarios. Further aspects of the energisation of electrons at collisionless shocks are discussed in the article by \citet{chapter6}.

Plasma turbulence generates inhomogeneous and intermittent patches of coherent structures, including eddies, flux ropes, and plasmoids separated by intense current sheets and shear layers, across the entire turbulent spectrum \citep{brunetti2007compressible, bruno13, matthaeus2015intermittency}. These structures mediate particle transport and energisation intermittently, for example, through magnetic reconnection. Under certain conditions, particles are trapped within these coherent structures. These trapped particles are inhibited in their  diffusion and experience intense perpendicular energisation in association with the merging or contraction of islands and plasmoids \citep{ambrosiano1988test, dmitruk2003test, drake2006electron, kowal2011magnetohydrodynamic, leroux2015kinetic,lemoine2021particle, lemoine2022first, lemoine2025effective, servidio2016explosive, lemoine2019generalized, trotta2020fast,  pezzi2022relativistic}. 

\begin{figure}[!htb]
    \centering   \includegraphics[width=\textwidth]{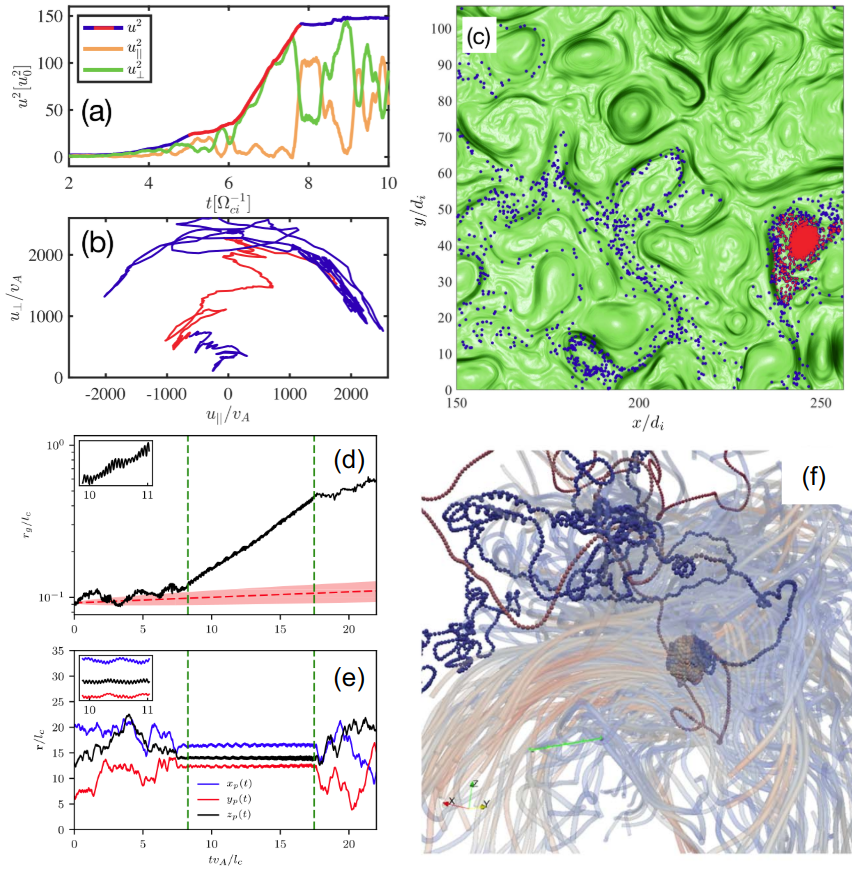}
    \caption{Role of particle trapping for particle energisation  in coherent structures. Left: thermal electrons embedded in plasma turbulence modelled with hybrid particle-in-cell simulations  \citep{trotta2020fast}. Right:  energetic protons in incompressible MHD turbulence \citep{pezzi2022relativistic}. In both cases, strong and preferentially-perpendicular energisation is associated with particle trapping in large-scale coherent structures. Trapping phases (red curves in panels (a) and (b), red points in panel (c), and the area  between the two vertical green lines in panels (d) and (e)) are preceded and followed by erratic particle motion in the entire computational domain across which particles undergo second-order diffusion in energy as well as isotropisation}
      \label{fig:TrottaPezzi}
\end{figure}

Figure~\ref{fig:TrottaPezzi} illustrates the role of particle trapping in coherent structures as an efficient particle energisation mechanism. 
The panels on the left-hand side show the results of a two-dimensional hybrid particle-in-cell simulation that investigates the energisation of trans-relativistic test-electrons with gyro-radii of order the ion inertial length  \citep{trotta2020fast}. 
The panels on the right-hand side show the results of an MHD simulation focused on fully-relativistic test-protons with gyro-radii in the inertial range of incompressible MHD turbulence \citep{pezzi2022relativistic}.
The electron energy evolution in panel (a) and the typical velocity-space trajectory in panel (b) indicate that the trapped electrons experience an intense perpendicular energisation during the trapping phase, which is marked with a red curve in panel (a) and red points in panel (c). After they escape from the trapping structures, the electrons cease their energisation and diffuse on constant-energy shells as seen in panel (b). Energetic protons follow a similar evolution. During their erratic motion in the turbulent electromagnetic field, a few protons experience a short period -- between the two green vertical lines in panels (d) and (e) -- in which they are trapped within a flux-tube-like structure shown in panel (f). During the trapping phase, their energisation is exponential as seen in panel (d) and mostly perpendicular to the magnetic field.

These simulations suggest that particle energisation in a turbulent environment is a multi-stage process in which particle trapping favours a strong energisation across the background magnetic field. This process is mediated by particle trapping and acts in conjunction with other energisation mechanisms, including collisionless shocks and magnetic reconnection \citep{giacalone2008energy, karimabadi2014link, zank2015diffusive, nakanotani2021interaction, trotta2020fast, trotta2021phasespace, trotta2022transmission, trotta20233D}, leading to  efficient acceleration of particles in space, astrophysical, and laboratory plasmas. 

In the solar corona, which serves as a testbed for particle acceleration, turbulent scattering leads to diffusive transport of flare-accelerated electrons, trapping them in the corona \citep{2018ApJ...862..158E}. For example, combined X-ray ($<100$~keV) and microwave ($>100$~keV) observations suggest the presence of scattering, leading to an energy-dependent relationship for the turbulence scattering mean free path such that the mean free path is inversely proportional to energy \citep{2018A&A...610A...6M}. Moreover, multiple studies of electron properties, determined from X-ray spectroscopy and imaging \citep{2011SSRv..159..301K,2017LRSP...14....2B}, suggest some form of coronal electron trapping. Observational evidence of this trapping includes spectral-index differences  \citep{2012ApJ...748...33C} and ratios of electron acceleration rates \citep{2013A&A...551A.135S}. If flare electrons stream from assumed acceleration sites in the tenuous corona to a dense \emph{thick-target} chromosphere \citep{1971SoPh...18..489B},  the spectra in the corona and chromosphere are expected to show a spectral-index difference of $\sim2$ based on the comparison of predictions for thin-target and thick-target bremsstrahlung in these regimes. However, observed differences are often less than 2, possibly indicating electron trapping \citep{2012ApJ...748...33C}, while the electron trapping is more efficient for the highest-energy electrons \citep{2018A&A...610A...6M}. Electron acceleration rates in the corona compared to the chromosphere indicate that some portion must be trapped in the corona \citep{2013A&A...551A.135S}. 

In the last few years, after the launch of  Solar Orbiter \citep{2020A&A...642A...1M}, stereoscopic flare observations, which view the flare from two different angles, are now possible. This new capability brings multiple opportunities for studies related to flare turbulence and particle acceleration. The Spectrometer/Telescope for Imaging X-rays \citep[STIX;][]{2020A&A...642A..15K} on Solar Orbiter, alongside Earth-orbiting X-ray observatories such as the Fermi Gamma-ray Burst Monitor \citep[GBM;][]{2009ApJ...702..791M} and the Advanced Space-based Solar Telescope Hard X-ray Imager  \citep[ASO-S/HXI; ][]{2019RAA....19..160Z},
observe flare X-ray emission from different locations. In particular, this allows us to probe the angular distributions of flare-accelerated electrons often known as the \emph{electron directivity}. This parameter is a key link to the underlying acceleration and transport processes, and hence, to the role of turbulence and particle diffusion \citep{2024ApJ...964..145J}.

\subsection{Spatial diffusion}\label{sect:spatial}

As discussed in the previous sections, thermal electrons in non-relativistic space and astrophysical plasmas follow the magnetic-field lines due to their small gyro-radius most of the time. Thus, in the classical case described by Braginskii \citep{braginskii1965transport}, perpendicular diffusion is weak \citep[see also][]{chapter7}. At small spatial scales $\sim \rho_{\mathrm{e}}$, the energy of the turbulent fluctuations, either pre-existing or self-generated through instability, is typically much less than the turbulent energy at large injection scales $\sim \ell_{\mathrm c}$ (see Sect.~\ref{turb_cascade}). However, there is growing evidence that also magnetic fluctuations at small scales suppress parallel thermal diffusion as shown by theoretical and numerical studies \citep{levinson92,pistinner1998self, 2018ApJ...867..154R, Beckman2022, yerger25} and in measurements in the solar wind \citep{gary99, halekas21, Coburn2024}; see also figure~1 of \citet{chapter7} and the detailed discussion therein.

The spatial transport of energetic electrons is very different from that of thermal electrons due to their large gyro-radii. If their energy is sufficiently large (i.e., $v_{\perp}\sim v\sim c$), their cyclotron-resonant wavenumber according to Eq.~(\ref{rescond1}),
\begin{equation}
k_{\mathrm{e,res}}\sim (\omega_{\mathrm r}-n\Omega_{\mathrm e})/c\sim n/\rho_{\mathrm e},
\end{equation}
eventually falls within the inertial range of the plasma turbulence. Galactic cosmic-ray electrons that escape supernova remnants and propagate through the interstellar medium are such a population of energetic particles. As seen in figure~4 of \citet{adriani2023direct},  cosmic-ray electrons have energies up to (at least) $\sim 1$\,TeV, although the highest energy of cosmic-ray electrons is still debated \citep{sudoh2023end}. Moreover, electron energies are significantly altered by radiative losses in contrast to protons \citep{2021A&A...650A..62C, evoli2021galactic,dorner2023cosmicray}. Considering the fiducial values of $1\,\mu$G for the magnetic field and $1$\,cm$^{-3}$ for the particle density in the interstellar medium, the gyro-radius $\rho_{\mathrm e}$ of a TeV electron is about $\sim 10^{-3}$\,pc and thus much smaller than the correlation length $\ell_{\mathrm c}\sim 10$\,pc of the turbulence but greater than the proton inertial length $d_{\mathrm p}\sim 10^{-6}$\,pc. Therefore,  the resonant wavenumber of these energetic electrons likely falls in the inertial range of the turbulence in the interstellar medium. At these energies, the energetic electrons diffuse similarly to protons, and the considerations about proton diffusion across the magnetic field also apply to electrons.  

In addition to the resonant diffusion mechanisms that depend on the available energy at the scale $k_{\mathrm{e, res}} \sim 1/\rho_{\mathrm e}$ \citep{shalchi2009nonlinear}, electrons may diffuse due to alternative mechanisms. An efficient particle diffusion mechanism is the large-scale field-line random walk (FLRW) of magnetic-field lines that induces particle diffusion across the background magnetic field \citep{laitinen16}. This diffusion occurs either homogeneously or ordered on large scales $\sim \ell_{\mathrm c}$. 
Turbulence is capable of suppressing thermal conduction by FLRW to values that are about two orders of magnitude below the Spitzer value in galaxy-cluster flows \citep{chandran1998thermal}. 
This behaviour is also observed in the case of super-Alfv\'enic turbulence at the outer scales, i.e., in the case of velocity perturbations at the injection scales that are greater than the Alfv\'en speed \citep{lazarian2006enhancement}. However, turbulence can also enhance heat conduction, depending on the plasma magnetisation and the turbulence driving \citep{lazarian2006enhancement}, in which case turbulence induces electron advection that provides effective heat diffusivity up to values exceeding the Spitzer value.

Understanding the properties of spatial diffusion  perpendicular to the magnetic field is a major challenge that still motivates significant theoretical \citep{matthaeus2003nonlinear,shalchi2004nonlinear,shalchi2009nonlinear,shalchi2010unified,shalchi2015perpendicular,shalchi2019heuristic,shalchi2021perpendicular} and numerical research efforts \citep{casse2001transport,shalchi2004nonlinear,demarco2007numerical,cohet2016cosmic,pucci2016energetic,arendt2020detailed,dundovic2020novel,mertsch2020test,reichherzer2022anisotropic,maiti2022cosmicray, kuhlen2025field}. 
From the theoretical perspective, different theories and models, including classical quasilinear theory and second-order quasilinear theory, are proposed for the calculation of the diffusion coefficient for the resonant diffusion parallel to the magnetic field. The assumption of a particular type of turbulence (composite slab and 2D, three-dimensional isotropic, critically balanced, etc.) allows us to evaluate the pitch-angle diffusion coefficient $D_{\xi \xi}$ \citep{shalchi2009nonlinear,schlickeiser2002cosmic}, which in turn depends on the resonances with the magnetic-field fluctuations at wavenumber $k_{\mathrm{e,res}}\sim 1/\rho_{\mathrm{e}}$. 
The parallel diffusion coefficient is then directly given by \citep{schlickeiser2002cosmic}
\begin{equation}
D_\parallel = \frac{3 v}{8} \int\limits_{-1}^{+1}  \frac{(1-\xi^2)^2}{D_{\xi \xi}}\,\mathrm d\xi,
\end{equation}
where $\xi$ is the pitch-angle cosine $\xi=\cos \theta = v_\parallel/v$.

\begin{figure}[ht]
    \centering   \includegraphics[width=\textwidth]{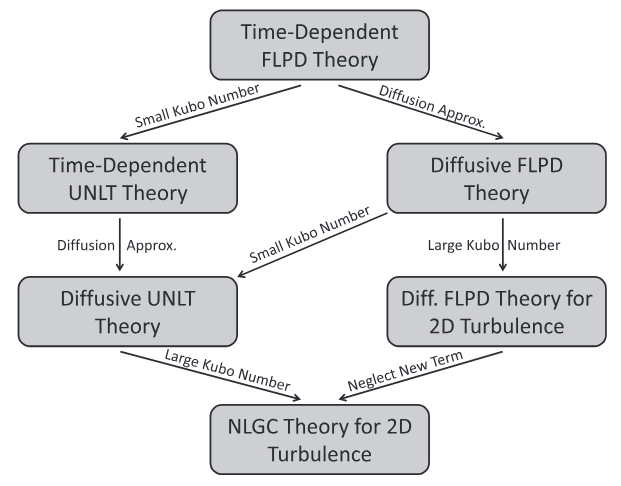}
    \caption{Schematic view of the links between a set of recent theories on perpendicular particle transport. The Kubo number $K = (\ell_{\mathrm c,\parallel}/\ell_{\mathrm c,\perp}) (\delta B_{\perp}/B_0)$ quantifies the intensity of the turbulent fluctuations modulated with the ratio of the parallel to the perpendicular correlation lengths $\ell_{\mathrm c,\parallel}$ and $\ell_{\mathrm c,\perp}$. FLPD: field-line--particle decorrelation, UNLT: unified nonlinear theory, NLGC: nonlinear guiding centre. The diffusive FLPD theory for 2D turbulence includes a new term that contains explicitly the field-line diffusion coefficient \citep{shalchi2021perpendicular}. If this term is neglected, the diffusive FLPD theory for 2D turbulence transitions into the NLGC theory for 2D turbulence. From \citet{shalchi2021perpendicular}}
    \label{fig:Shalchi2021}
\end{figure}

The parallel diffusion coefficient is one of the key ingredients for the calculation of the perpendicular diffusion coefficient $D_\perp$, and different theories and models exist for the evaluation of $D_\perp$. These include the nonlinear guiding-centre \citep[NLGC;][]{matthaeus2003nonlinear}, the unified nonlinear \citep[UNL;][]{shalchi2010unified}, and the field-line--particle decorrelation \citep[FLPD;][]{shalchi2021perpendicular} theories.
Figure~\ref{fig:Shalchi2021} illustrates the links and inter-dependencies between these theoretical frameworks. Most of them rely on a similar methodology which we briefly sketch out below. The goal of all these theories is the calculation of the perpendicular diffusion coefficient $D_{\perp}$ as defined in the Taylor--Green--Kubo (TGK) formulation \citep{kubo1957statistical}:
\begin{equation}
D_{\perp} = \int\limits _0^\infty  \langle v_x(t) v_x(0)\rangle \,\mathrm dt,
\end{equation}
where $\langle \dots \rangle$ denotes the ensemble average, and we assume that the background field $\vec B_0$ is along the $z$-direction. The derivation procedure is then as follows:
\begin{enumerate}
    \item The equations of motion are solved in a given magnetic-field configuration to calculate the second-order perpendicular correlation $\langle v_x(t) v_x(0)\rangle$, usually with the additional assumption that the magnetic-field perturbations are slow and that their component parallel to $\vec B_0$ is negligible ($\delta B_z=0$). The second-order perpendicular correlation $\langle v_x(t) v_x(0)\rangle$ can be written as a fourth-order correlation including the parallel speed of the gyro-centre ${\tilde v}_z$ and the perpendicular magnetic-field perturbations $\delta B_{\perp}$.
    \item The fourth-order correlation is separated into two second-order auto-correlations: one for the parallel velocity and one for the perpendicular magnetic-field perturbations.
    \item A description for the auto-correlation function of the parallel velocity is then introduced, e.g., an exponential decay. This description relates the auto-correlation function of the  parallel velocity with the parallel diffusion coefficient $D_{\parallel}$.
    \item  Corrsin's hypothesis for homogeneous turbulence is usually invoked to relate the auto-correlation function of the perpendicular magnetic field to the energy spectrum of the turbulence \citep{shlien74,tautz10}.
    \item Finally, ensemble averages are calculated by assuming some distribution of particles and a closure for the particle mean displacements (e.g., a diffusive closure).
\end{enumerate}

These steps generally lead to a nonlinear integral equation for $D_\perp$ that depends on $D_{\parallel}$ and on the type, intensity, and spectral properties of the magnetic-field turbulence. The above-listed theories have been tested against numerical simulations of charged test particles in turbulent fields, generated either synthetically \citep{dundovic2020novel} or through MHD simulations \citep{cohet2016cosmic, maiti2022cosmicray}. A still-debated aspect concerns whether the perpendicular diffusion coefficient has the same energy dependence as the parallel diffusion coefficient for $\rho_{\mathrm{e}}<\ell_{\mathrm c}$. Indeed, different models predict that the parallel and perpendicular diffusion coefficients should scale similarly with particle energy $W$; however, there is growing evidence for different energy dependencies of $D_\parallel$ and $D_\perp$ in numerical simulations \citep{demarco2007numerical, dundovic2020novel, kuhlen2025field}.

An additional fundamental problem is related to the nature of the mechanisms responsible for parallel diffusion. Resonant diffusion critically depends on the power in the fluctuations at the resonant scale $k_{\mathrm{e,res}}$. Diffusion is weakened when the power is weak or absent at this scale, as it is the case for electrons with low to intermediate energies in anisotropic Alfv\'enic or slow-magnetosonic turbulence. Fast-magnetosonic turbulence can partially overcome this issue since it cascades isotropically \citep{cho2002compressible}. However, the spectrum of fast-magnetosonic turbulence is affected by various damping mechanisms depending on the properties of the background medium, such as the plasma-$\beta$. 

Other processes that contribute to particle diffusion when resonant diffusion is weak include FLRW \citep{pezzi2024galactic} and small-scale magnetic-field curvature \citep{kempski2023cosmicray,lemoine2023particle,kempski2025}. The FLRW-driven diffusion provides an energy-independent diffusion coefficient which adds to the classical resonant diffusion for galactic cosmic rays \citep{pezzi2024galactic}. Curvature-driven diffusion instead provides an energy-dependent diffusion coefficient with a scaling that is similar to that predicted by standard resonant diffusion. Further efforts are still needed to explore the importance of these mechanisms depending on the key parameters of the turbulence (e.g., $\delta B/B_0$ or the level of intermittency) and to understand whether they induce features in the cosmic-ray energy spectrum. 

The  microphysics of energetic-particle transport has vast applications in astrophysics. The perpendicular and parallel diffusion coefficients serve as inputs for the solution of advection-diffusion transport equations, such as the Parker equation and the focused-transport equation \citep[for a review, see][]{zank2014transport}. These equations describe particle transport in various contexts, including the galactic-cosmic-ray transport in the interstellar medium \citep{skilling1971cosmic, schlickeiser2002cosmic, blasi2013origin, dorner2023cosmicray} and the energetic-particle transport in the heliosphere \citep{parker1965passage, gordovskyy2014particle, leroux2015kinetic, zimbardo2017fractional, pezzi2021current, engelbrecht2022theory, wijsen2022observations}. 

\section{Conclusions}\label{sect:conclusions}

The scales associated with kinetic electron physics are typically the smallest characteristic scales associated with collective behaviour in plasmas. The electron gyro-radius $\rho_{\mathrm e}$ is often much smaller than the scales associated with variations in the magnetic field, even when the field is highly tangled. In these cases, the electrons follow the magnetic-field lines, which allows us to understand and use them as tracers of magnetic connectivity in plasma systems. Since electron heat transfer is strongest along the magnetic field, also heat transfer does not follow straight lines in these tangled field configurations. 

Tangled magnetic fields can lead to drifts, trapping, and the scattering of electrons. Trapping occurs in localised regions of low field magnitude when particles experience consecutive reflections due to the mirror force. There are two main pathways for electrons to experience scattering: (a) via resonant wave--particle interactions when electrons experience acceleration through quasi-steady wave electric fields in their own reference frame; or (b) when the gyro-radii of the electrons are comparable to the scale of variations in the magnetic field, leading to spatial diffusion along and across the field.

Trapping, wave--particle interactions, and cross-field diffusion are often related and mutually depend on each other. Cosmic magnetic fields are almost always inhomogeneous and tangled, from the phase of magnetogenesis to the turbulent evolution of evolved plasmas. Therefore, electron transport is generally a result of a complex interplay of streaming along tangled field lines, gyro-centre drifts, collisions, trapping in inhomogeneous structures, and scattering on electric and magnetic fluctuations. All of these processes impact the large-scale behaviour of the plasma, for instance, through direct modifications of the electron fluid moments, inter-species coupling, and anomalous resistivity.

In-situ observations in space plasmas open a unique window into the understanding of this complex interplay between large-scale and small-scale processes through the simultaneous multi-scale measurement of electromagnetic fields and electron velocity distribution functions. 
With the help of stereoscopic and polarisation observations from Solar Orbiter and PADRE, we are now able to infer the electron pitch-angle distribution in solar flares. These observations provide a new diagnostic for the study of electron transport, the presence and nature of turbulence, and the structure of tangled magnetic fields in the flaring corona and similar cosmic plasma environments. Extrapolation from these heliospheric plasmas provides us with the opportunity to understand the transport of electrons in tangled fields across the Universe. As a first step and in dialogue with astrophysical observations, we must understand how tangled fields are in different cosmic plasmas beyond the heliosphere. By extrapolation from our knowledge about heliospheric plasmas, we must then quantify the relative importance of trapping, wave--particle interactions, and diffusion in these environments.

The inclusion of complex electron transport in global plasma models is a central challenge for plasma theory. For instance, simulations of magnetogenesis, cosmic dynamos, and the tangling of the magnetic field must account for electron-kinetic effects self-consistently. As another example, electron heat conduction plays an important role in the magnetothermal instability. If heat flux is suppressed, for example, by resonant wave--particle interactions, this instability is quenched, with profound implications for cosmic dynamos. 
Simulations also show that particle energisation in turbulent fields is a multi-stage process that involves electron-kinetic physics on all relevant scales. Trapping in islands and other localised field structures has emerged as an efficient energisation process. However, it is still not understood how this process compares to other energisation mechanisms. The answer to this fundamental problem involves not just the characterisation of the nature and occurrence of trapping structures. It also requires a careful analysis of the interactions responsible for the trapping and de-trapping of  electrons as well as the involved feedback loops between kinetic-scale and global processes.

Electron transport in  tangled magnetic fields is an important contemporary science challenge for the fields of heliophysics and astrophysics. Due to the complexities associated with the self-consistent interdependencies between global and kinetic-scale processes in this context, a truly multi-disciplinary approach of combining in-situ plasma measurements, remote-sensing plasma observations, and plasma theory and simulations is required. 
This approach must bridge from detailed in-situ measurements of the smallest characteristic plasma scales to the understanding of the largest plasma systems in the Universe.

\backmatter

\section*{Acknowledgements}
All authors thank the International Space Science Institute (ISSI) for
hosting the workshop on ``Electron Kinetic Physics: The Next Frontier in
Space and Astrophysical Plasmas'' (22--26 April 2024). D.V.~is supported by STFC Consolidated Grant ST/W001004/1. I.S.~is supported by the Swedish Research Council Grant 2016‐0550, the Swedish National Space Agency Grant 158/16, and the Knut and Alice Wallenberg foundation. O.P.~acknowledges the support of the PRIN 2022 project ``The ULtimate fate of TuRbulence from space to laboratory plAsmas (ULTRA)'' (2022KL38BK, Master CUP: B53D23004850006) by the Italian Ministry of University and Research, funded under the National Recovery and Resilience Plan (NRRP), Mission 4 – Component C2 – Investment 1.1, ``Fondo per il Programma Nazionale di Ricerca e Progetti di Rilevante Interesse Nazionale (PRIN~2022)'' (PE9) by the European Union – NextGenerationEU.
N.J.~gratefully acknowledges financial support from the Science and Technology Facilities Council (STFC; grants ST/V000764/1 and ST/X001008/1). A.A.~acknowledges the support of NASA contract NAS5-02099 for the use of data from the THEMIS Mission, NASA's CubeSat Launch Initiative for ELFIN's successful launch, and critical contributions by numerous ELFIN team members supported by NASA grant 80NSSC22K1005 and NSF grants AGS-1242918 and AGS-2019950. M.R.~gratefully acknowledges support from the ANID-FONDECYT grants 1191673 and 1260829, as well as from the Center for Astrophysics and Associated Technologies (CATA; ANID Basal grant FB210003). 

\section*{Declarations}
\subsection*{Conflict of interests}
Not applicable

\bibliography{entangled_fields}

\end{document}